# Astronomy's relationship with the lands and communities of Maunakea

Jean-Gabriel Cuby*[a], Christine Matsuda[b], Rich Matsuda[c], Andy Adamson[d], John O'Meara[c], Nadine Manset[a]

[a]Canada-France-Hawai'i Telescope, 65-1238 Mamalahoa Hwy, Waimea, HI 96743, [b]Maunakea Observatories, Maunakea, Island of Hawai'i, [c]W. M. Keck Observatory, 65-1120 Mamalahoa Hwy, Waimea, HI 96743, [d] The Gemini International Observatory, operated by NSF NOIRLab, 670 N Aohoku Pl, Hilo, HI 96720

**ABSTRACT**

Astronomy is at a turning point in its history and in its relations with the Indigenous peoples who are the generational stewards of land where several of our main observatories are located. The controversy regarding the further development of astronomy facilities on Maunakea is probably the most significant and publicized conflict about the use of such land in the name of science. Thousands have stood in resistance, elders were arrested, and the community is divided. Astronomy's access to one of its most emblematic sites is at risk. This situation challenges our professional practice, the projects we build on Indigenous lands, and our relationships with the people who live within these lands and with society in general. This paper attempts to share the perspective of the authors on the historical events, including the very recent past, through the lens of our understanding and opinions; to provide transparency, with humility, into our process of introspection and transformation; and to share our hopes and ambitions as leaders from Maunakea Observatories for the future of astronomy in Hawai'i, as advocated by the Astro2020 report from the U.S. National Academies of Sciences, Engineering, and Medicine; and to suggest ways for the profession to commit to this long-term vision.

**Keywords:** Maunakea, Culture, Community, Protests, Activism, Mutuality, Stewardship, Governance, Community Astronomy Model

## 1. INTRODUCTION

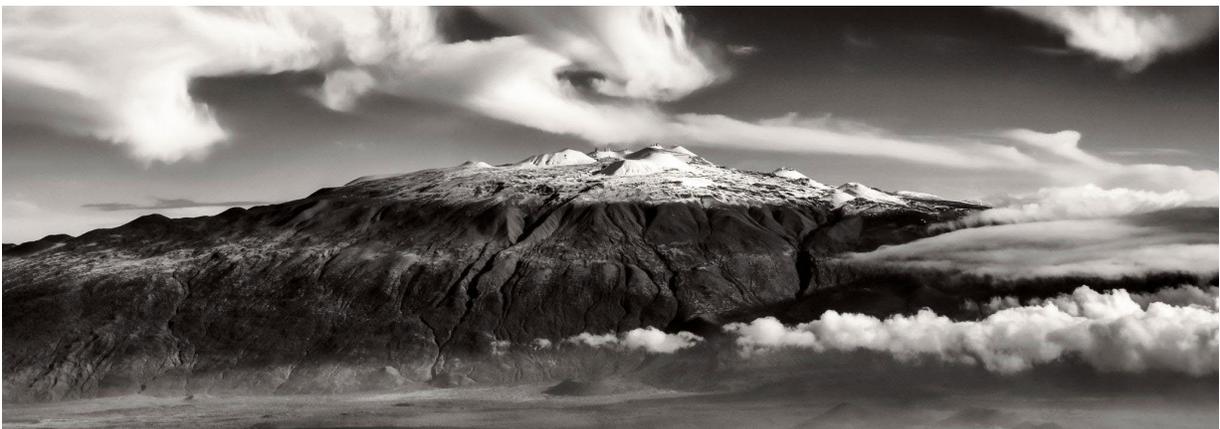

Figure 1. Maunakea, Hawai'i Island. Maunakea rises from the seafloor to an elevation of 13,803 feet (4,207 meters), making it the highest mountain in the Pacific, and, if measured from its base, the tallest mountain in the world. Photo credit: Ethan Tweedie

*cuby@cfht.hawaii.edu

**The relationship between astronomy and Maunakea[1], a site of cultural, historical, and scientific significance**

Astronomy is often considered the oldest of the sciences. For millennia, before humankind had access to space, astronomy relied on observing the sky through the atmosphere, from the ground. The UNESCO-IAU Portal to the Heritage of Astronomy lists astronomy-related artifacts, buildings, and monuments dating back to the Neolithic period. The search for suitable sites for observing the sky, or for the construction of such monuments, probably began as early as this period. Throughout our history as people, across geographies and societies, our curiosity about the universe beyond has fueled the development of a rich diversity of scientific disciplines across cultures. In Polynesia and Micronesia, Native scientists built generations of observation-based knowledge into complex systems: a treasured tradition of celestial navigation that voyagers used to navigate the vast Pacific. Their students were studying the stars from astronomy and navigational heiau[2] at the same time that civilizations on the other side of the world were experimenting with parallel observation technologies and the construction of early telescope instruments.

By the second half of the 20th century, observatories were gradually moved to the outskirts of cities to escape light pollution and, as means of transport improved, to nearby mountains to offer better atmospheric conditions. After the air transport revolution of the 1960s, the search for astronomical sites took on a new dimension and became borderless. This marked the beginning of the boom in astronomical sites in Chile and Hawaiʻi. The history of the development of astronomy facilities in Hawaiʻi is well documented, for example in Swanner [1], Marichalar [2], Konchady [3], and Avallone[3] [4]. Today, and throughout their operation on Maunakea, the Maunakea Observatories together constitute the most scientifically productive observatory complex in the world.

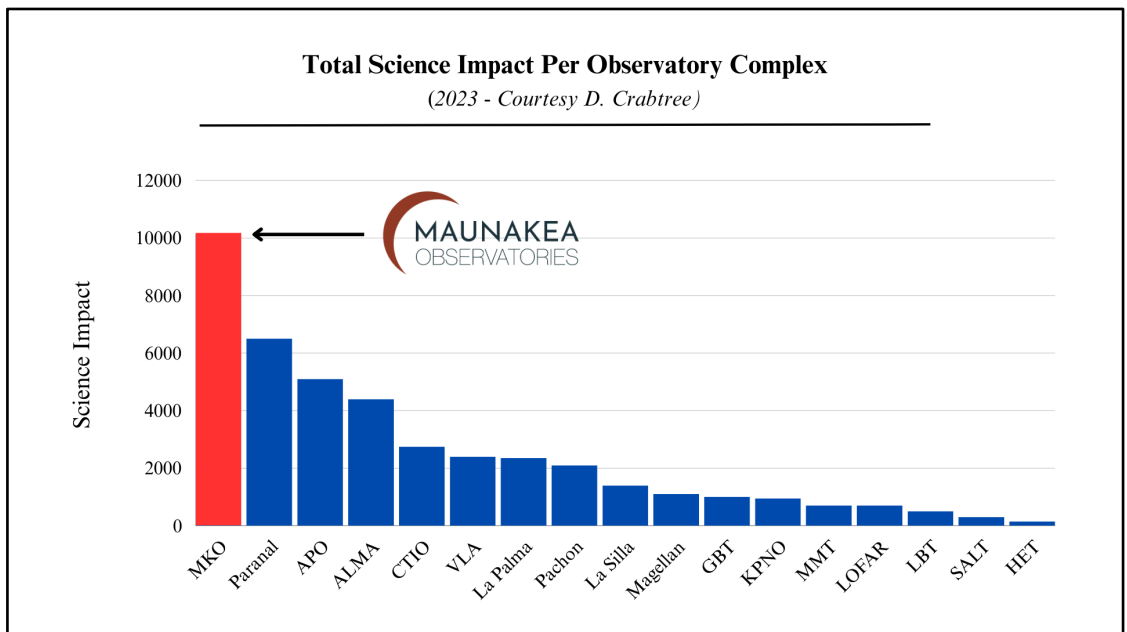

Figure 2. Chart shows the total scientific impact of major observatories around the world. The impact of a paper is defined as the number of citations to the paper divided by the number of citations received by the median ApJ paper of the same year [5].

Astronomy as we practice it today is a technology-enabled science. The profession recognizes that astronomical knowledge depends as much on the engineers and technicians who build and operate telescopes and instruments as on the astronomers who analyze and interpret the data within a theoretical and numerical framework. We argue that recognition, much greater than has been shown historically, is owed to the lands and peoples on which our observatories are located, without which our astronomical knowledge would be far more limited.

**Indigenous lands and sacred mountains**

It is not the intention of this paper, nor the competence of its authors, to define what Indigenous lands are. Suffice it to say that many astronomical observatories are located in places where Indigenous peoples have stewarded lands and held spaces as sacred for many, many generations.[4] The history of these lands and astronomy's presence is different at each site, and we discuss the particular case of Maunakea in Hawaiʻi in this paper.

Maunakea, like many other mountains around the world, has great cultural significance and is often referred to as a sacred site by the Hawaiian people. "Sacred" in this usage is often misunderstood by many in the profession. The meaning of the word in the Western world is very different from its meaning in many Indigenous worldviews where land, people, natural elements, living things, and language are interconnected. In many Indigenous cultures, a bird, a stone, a flower, a river, water, fire, or a mountain can be sacred. "To the Western ear, "sacred" may be synonymous with "sacrosanct" — inviolably holy — but to an Indigenous culture, a place labeled as "sacred" may instead mean something spiritually alive, culturally essential, or simply deserving of respect," [6].

Mountains have probably always inspired and fascinated people. Many mountain tops are emblematic in their countries and around the world, from Mount Everest to Mount Fuji, Kilimanjaro, Aconcagua, Denali, Mont Blanc, and many more. Would the populations close to those mountains authorize the construction of multi-story buildings on the summit of these iconic mountains? For most of them, probably not. Maunakea is the highest mountain in the Pacific, and along with Maunaloa, has been an iconic mountain for the Polynesian voyagers and Hawaiian people for centuries. However, 13 telescopes have been built on its summit, many of them highly visible from anywhere with a view of Maunakea's summit.[5]

**Sharing and learning within communities**

What would be the state of our astronomical knowledge if we didn't have observatories in Hawaiʻi, Chile, the United States, Australia, or elsewhere, where Indigenous peoples live? Would we know as much about exoplanets, dark energy, or the black holes at the center of the Milky Way or M87? Certainly not.

How is this new scientific knowledge passed on, shared, and benefited from the local communities on whose land our observatories are located? How, in reverse, is Indigenous knowledge transmitted, shared, and benefited from the astronomical communities? This question of reciprocity between astronomical knowledge and Indigenous knowledge is at the very heart of our relationship with the Indigenous communities with whom we interact. It is also at the heart of past and present controversies between astronomy and local populations.

Released in 2021, the US Decadal Survey on Astronomy and Astrophysics 2020 (Astro2020) included a formal Panel on the State of the Profession and Societal Impacts for the first time [7]. The panel's report is included in Appendix N of the Astro2020 report, and the main report includes an entire chapter (Chapter 3 entitled "The Profession and Its Societal Impacts: Gateways to Science, Pathways to Diversity, Equity, and Sustainability") that contains strong recommendations on how the professions should engage differently with local populations where we practice our profession. The recommendation for developing a **community astronomy** model is particularly significant. In many fields of science, there is a growing recognition of the need to strengthen local communities and integrate local knowledge into the traditional Western approach to science that has dominated recent centuries. It is telling, for example, that Wikipedia has entries for community archaeology or community forestry, testifying to the fact that these concepts exist and are well-developed. We develop our vision of what community astronomy might look like below, recognizing and making clear that any such model could only be successful if developed in complete collaboration (co-development) with the local people involved.

In this paper, we aim to share the perspective of the authors on the historical events, including the very recent past, through the lens of our understanding and opinions; to provide transparency, with humility, into our process of introspection and transformation; and to share our hopes and ambitions for the future of astronomy in Hawaiʻi, as advocated by the Astro2020 report from the U.S. National Academies of Sciences, Engineering, and Medicine; and to suggest ways for the profession to commit to this long-term vision. This paper presents a view that is, by its nature, limited in breadth and not universally-shared; we are institutional leaders who work with the Maunakea Observatories, and this is a personal perspective. Some of the authors were born and raised in Hawaiʻi and some have only recently adopted Hawaiʻi as home; none of us claim Native Hawaiian ancestry, nor do we claim to have an experts' understanding in the history and culture of the Hawaiian people.

Astronomy in Hawaiʻi is in the midst of what we as authors hope will be a profound and durable transformation to the ways we engage as members of our community, and to the ways we practice astronomy as institutions. In order to contextualize the gravity and necessity of this fundamental shift, we begin by looking back at the history of Maunakea land tenure—its complexities, including both hopeful reference points and profound injustices—and that history's implications for the present day. We discuss how and why the new governance model, overseen by the newly-established Mauna Kea Stewardship and Oversight Authority, was formed, and what it will mean for Hawaiʻi astronomy moving forward. We end with a look ahead to the vision hoped for by the authors, including the efforts currently underway, an early articulation of our efforts to create that fruitful future.

## 2. MAUNAKEA, HAWAIʻI

**Maunakea, Hawaiʻi: A complex story**

> *The tangible and intangible resources that contribute to the significance and integrity of Mauna Kea…are critical to the history and evolution of Kānaka ʻŌiwi culture and people; the connection between Kānaka ʻŌiwi and the mountain cannot be severed because it is first and foremost a genealogical one - to deny this relationship is to deny the ancestry of a lāhui.* The Kaliʻuokapaʻakai Collective, National Register of Historic Places Registration Form, 2021 [8]

The discourse surrounding Maunakea issues often places judgment on the relative virtue of different groups' preferred uses of the site: the purity of scientific curiosity, or the authenticity of the cultural practice, for example. But what that kind of analysis fails to uncover is the larger set of issues looming behind those value judgments: Who has the privilege and responsibility— or the power—to decide how the land is used in the first place?

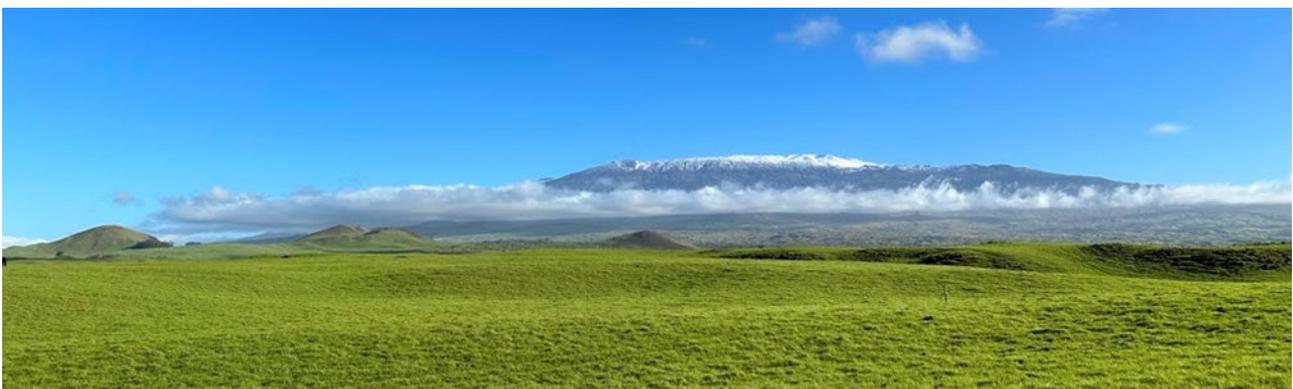

Figure 3. Maunakea, Hawaiʻi Island. Photo credit: Christine Matsuda

**Maunakea in the Kingdom of Hawaiʻi**

Cultural tradition and inherited knowledge inform us today that Maunakea has always been seen, from the days of the earliest established communities in Hawaiʻi, as a place of particular importance. Many ancient and contemporary moʻolelo

(history, chronicle, records), as well as traditional oli (chants) and mele (songs), document the sacred stature of the mauna.[6] In contemporary Hawai'i, its stature as an extraordinary place is understood through many perspectives—those of Hawaiian cultural practitioners, artists, hunters, astronomers, ecologists, archaeologists, and many more.[7]

The Hawaiian Islands were first settled as early as 400 C.E. by Polynesian deep-sea explorers using sophisticated systems of celestial navigation to travel throughout the vast Pacific [9]. Native Hawaiians thrived in highly developed, interconnected societies throughout the islands. The traditional land governance system in Hawai'i was based on principles of mutual, communal land tenure—no one person or family-owned any parcel of land, but rather had stewardship responsibilities they were privileged to fulfill [10]. This perspective—one commonly held by native and Indigenous peoples' traditions around the world—continues to inform cultural practice and personal responsibility today.

In 1848 the Hawaiian Kingdom[8] converted the land governance system to one of western-style private ownership referred to as the Māhele, a comprehensive division of Hawai'i's four million acres (16,200 km$^2$) into four categories of ownership.[9] From that point forward, lands could be granted, bought, or sold, but Maunakea never was. Maunakea remained under the control of the government of Hawai'i, under the jurisdiction of the king.

**Jurisdiction of Maunakea at the point of the Overthrow**

The next wave of change to the system of land governance in Hawai'i had profound and tragic repercussions that remain raw and painful to the present day: the unlawful overthrow of the Kingdom of Hawai'i by a group of American businessmen who preferred to avoid international tariffs by becoming part of the United States.[10]

In January 1893, American businessmen, who called themselves the Committee of Safety, proclaimed the establishment of a Provisional Government. The Provisional Government took control of all Crown, Government, and Public Lands - including Maunakea. Six months later, the Provisional Government declared itself to be the Republic of Hawai'i; then on January 24, 1895, while imprisoned in her palace, Queen Lili'uokalani was forced to officially abdicate her throne by representatives of the Republic of Hawai'i. Shortly after becoming president, McKinley signed the Newlands Joint Resolution—also referred to as the Joint Resolution of Annexation[11]—in which the self-declared Republic of Hawai'i ceded sovereignty over the Hawaiian Islands to the United States, despite the vehement and unyielding opposition of the Hawaiian people and the Queen.[12]

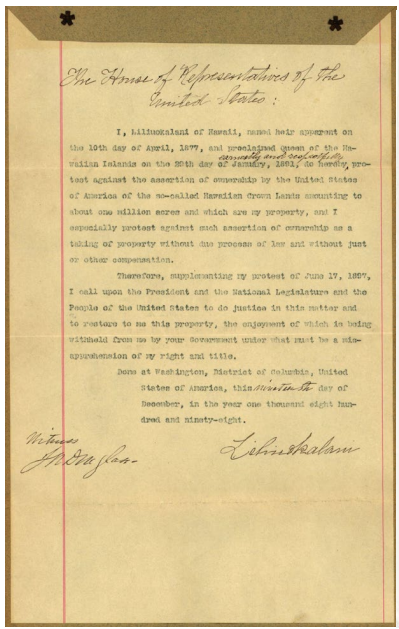
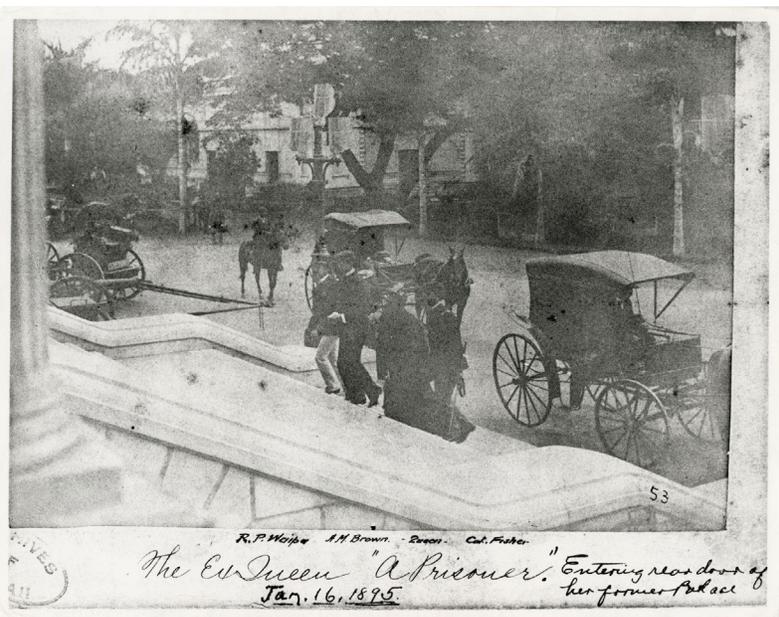

Figures 4 & 5. Left: [Letter from Liliʻuokalani, Queen of Hawaiʻi to U.S. House of Representatives protesting U.S. assertion of ownership of Hawaiʻi](#), December 19, 1898, National Archives; Right: [Queen Liliʻuokalani escorted into her palace as a prisoner](#), Jan 16, 1895, Hawaiʻi State Archives.

Along with the sovereignty of the nation, "…the Republic of Hawaiʻi also ceded 1,800,000 acres of crown, government, and public lands of the Kingdom of Hawaiʻi, without the consent of or compensation to the Native Hawaiian people of Hawaii or their sovereign government," [10]. That is how the United States government came to see itself as the owner of Hawaiʻi's 1.8 million acres (7,300 km$^2$) of lands and natural resources, including those atop Maunakea.

A hundred years later, in 1993, President Clinton signed the US Senate Joint Resolution 19, known as the Apology Resolution, which stated the following:

> *The Congress…apologizes to Native Hawaiians on behalf of the people of the United States for the overthrow of the Kingdom of Hawaii on January 17, 1893, with the participation of agents and citizens of the United States, and the deprivation of the rights of Native Hawaiians to self-determination.* [United States Congress Joint Resolution](#), 1993 [10]

**The 50$^{th}$ State and the Public Lands Trust**
The US Congress made Hawaiʻi a state in 1959 via the Admission Act.[13] In that act, Congress transferred 1.4 million of the ceded land acres,[14] (5,700 km$^2$) including Maunakea, to the newly-formed State government of Hawaiʻi. Section 5(f) of the Admission Act, directing the state to hold the lands in trust, listed the following five purposes, for the lands' use:

> *The lands granted to the State of Hawaiʻi by subsection (b) . . . shall be held by said State as a public trust for the support of the public schools and other public educational institutions, for the betterment of the conditions of native Hawaiians, as defined in the Hawaiian Homes Commission Act, 1920, as amended, for the development of farm and home ownership on as widespread a basis as possible[,] for the making of public improvements, and for the provision of lands for public use.* [An Act to Provide for the Admission of the State of Hawaiʻi into the Union](#) [11]

Known as the Public Land Trust, these lands include Maunakea—known to be of the utmost sensitivity and importance—seized unlawfully from the monarchy of the Hawaiian Kingdom, transferred to the United States by the self-proclaimed Republic of Hawaiʻi, and transferred in turn to the State of Hawaiʻi for stewardship, held by the State as a public trust. Under the State of Hawaiʻi, The Department of Land and Natural Resources was established[15] to manage, administer, and exercise control over public lands.

**Authorizing construction on the summit**
In the words of John Jefferies, the first director of the Institute for Astronomy (IfA) at the University of Hawaiʻi (UH) in his first-hand history of that era:

> *Following statehood in 1959, the desire for change from the old ways of Hawaii was widespread, and John Burns, as the new Governor, was its incarnation and powerful advocate. He believed that a strong University could become a major factor in shaping a new future for the State, and had found a leader after his own heart in Thomas Hamilton, whom he saw appointed to the University presidency.* John Jefferies, [Dawn of a Brilliant Opportunity](#) [12]

The 1960s were a period of great expansion for higher education in Hawaiʻi. The community college system was established in 1964, with four new campuses, and in the next few years, the University established three new schools at its flagship campus at UH Mānoa [13].

At the same time, the Island of Hawaiʻi was at an economic inflection point. The sugar industry was waning, and in 1960 a catastrophic tsunami devastated the island's business epicenter in Hilo. The executive secretary of the Hawaiʻi Island Chamber of Commerce, Mitsuo Akiyama, was charged with finding solutions on behalf of the Hilo business community. Understanding the potential of higher education and what university expansion could mean for Hilo, and recognizing the unique nature of Maunakea as a potential site for astronomical research, Akiyama wrote to universities in the U.S. and in Japan asking them to evaluate Maunakea as an observatory site. Gerard Kuiper[16] of the University of Arizona Lunar and Planetary Lab responded, already familiar with the Hawaiʻi Island mountains and their potential [14]. From John Jefferies' account:

> *Kuiper came to believe that Mauna Kea could well be an even better site and perhaps the matchless one that he had been seeking. To follow this up, he enlisted the support of the Hilo Chamber of Commerce–and particularly of its Secretary, Mitsuo Akiyama, to establish a presence in the city as a start of his pioneering study. At the same time, he used his ample powers of persuasion to convince Governor Burns to build a path to the summit so as to test its astronomical quality. This road, constructed in 1963, was nothing more than a rudimentary track, very rough and very steep, but one with which we were to become only too familiar! With access thus obtained, Kuiper, in early 1964, set up a small site-testing dome on Puu Poliahu[17] and sent one of his staff to obtain observations of the seeing during a period of about six months in 1963/4.*
> John Jefferies, Dawn of a Brilliant Opportunity [12]

With Jefferies' involvement, the University requested a lease from the Department of Land and Natural Resources (DLNR), which was granted in 1968. The purpose of executing the lease was clear to all parties—the goal was to operate the Maunakea Science Reserve as a scientific complex to establish astronomy in Hawaiʻi [Stewardship of Maunakea, University of Hawaiʻi, accessed May 2024]. The lease includes provisions that prohibit waste, protect objects of antiquity, prevent the introduction of invasive species, etc., but it is clear that the purpose of these lands is not for ecology, conservation, or cultural practice. The purpose is the construction of an astronomy complex:

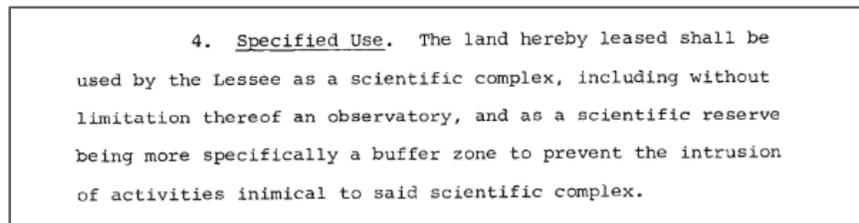

Figure 6. Excerpt from General Lease No. S-1491, executed between the State of Hawaiʻi and the University of Hawaiʻi for the use of Maunakea in 1968.

An important distinction—one that has been a sticking point in the local community—is what's missing in this lease document: there is no exchange of money. The lease is considered gratis, as a government (State of Hawaiʻi DLNR) to government (University of Hawaiʻi) agreement. Subsequently, when the University of Hawaiʻi eventually established subleases to the individual observatory nonprofits that would build on the summit, each sublease was also effectively gratis, in terms of monetary exchange at $1 per year.[18] The agreements are based on a model that is standard practice amongst universities and astronomy complexes,[19] but has been confusing to the community at large. The agreements guarantee observing time on each of the telescopes for the IfA, and over time, funding models have been established between the observatories and the State to support the maintenance of the astronomy complex's shared infrastructure, through subsequent financial agreements.

The Hawaiʻi State government identified the University of Hawaiʻi as the host for the State's astronomy program, which in turn led to the formation of the UH IfA [15]. The success of the IfA as a global leader in astronomy today is due in very large part to the university's access to Maunakea over the course of the last half-century. But with no "rent" revenue generated by the lease or subleases, there is no funding that flows to the Office of Hawaiian Affairs to be spent alongside other public trust revenues for the betterment of the Hawaiian people or the public at large.[20] That the primary "currency" of value at play in these agreements is telescope time is seen as acquisitive and unfair by beneficiaries of the Public Land Trust who are not affiliated with the observatories.

**Through the looking glass: Conflicting realities**
There are those in the community in Hawaiʻi—from scholars to academics, attorneys, activists, and cultural practitioners—whose perspective on this history demands that we not ignore the logical leaps that led the territorial and state governments to create the legal framework described in the sections preceding this one. Many challenge the validity of UH's lease to Maunakea lands based on what they point out as a fundamental fallacy: that the lands were legally held by the state in the first place.[21] Understanding that the legal reality of our present-day system recognizes these ceded lands as the responsibility of the State of Hawaiʻi, under current law the State is required to uphold and protect the rights customarily and traditionally exercised by Native Hawaiians [16]. The primary focus of the Maunakea lease as originally envisioned was only about the development of astronomical observatories, ignoring the fundamental fact that throughout the history of this place, Maunakea has always been a place of "cultural significance for Native Hawaiians, with resources and cultural sites essential to Native Hawaiian traditional and customary practices, specifically tied to Mauna Kea," [17].

This tension creates a juxtaposition of realities, each held tightly by the people who subscribe to each perspective. The first is one where a state has the clear legal right to allocate its own lands to its own university for educational purposes. That is set against a different, parallel reality where stolen land keeps changing hands, becoming further and further removed from its generational stewards and their ability to interact with a place deemed sacred. Both exist and are true, at the same time.

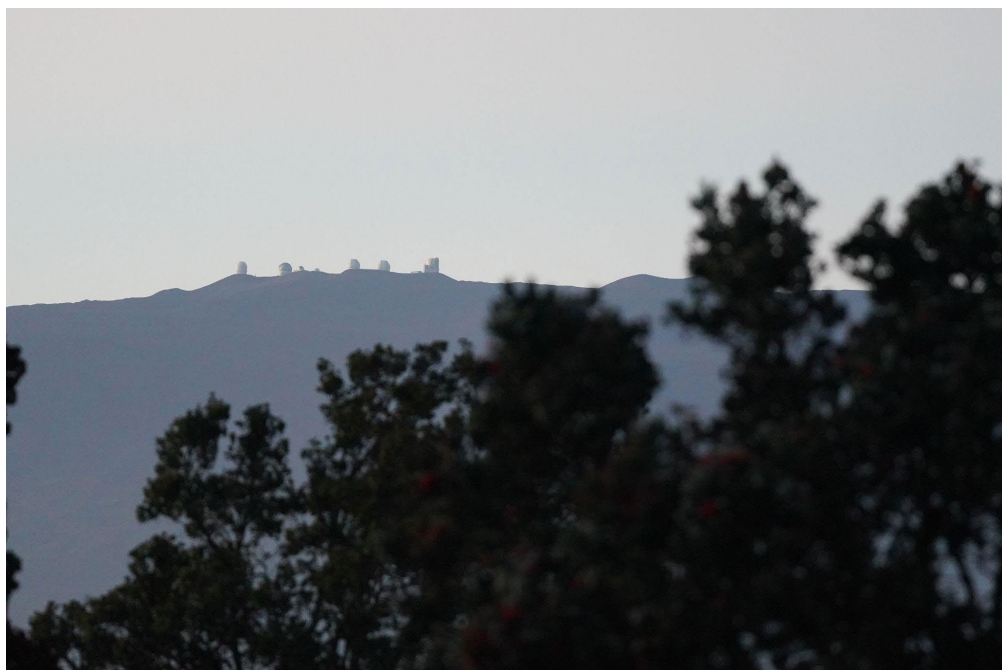

Figure 7. A view of Maunakea from the town of Waimea on Hawaiʻi Island. Photo credit: Cory Lum.

An area where one might hope to find common ground between these realities is in the overlap between two of the stated public land trust purposes: support for educational institutions, and the benefit of the Native Hawaiian people. As current IfA director, Doug Simons points out, "Observing time on the Maunakea Observatories is fundamental to the success of the State of Hawai'i's astronomy program. From that key resource stems educational opportunities, scientific prestige and leadership for Hawai'i, hundreds of millions of dollars in extramural funding, advanced technology development programs, broad economic benefit and diversification for our communities, and much more. The IfA's program helps leverage that resource for multilateral benefits, consistent with the mission of the University of Hawai'i, and the vision of those who inspired Maunakea astronomy decades ago," [15].

A thriving IfA—one of the most prestigious programs in the world—could have already become a place where Native Hawaiian astrophysicists would be publishing ground-breaking science every semester, if early prioritization and decision-making had been handled differently. One can look to the examples of the University's [medical](#) and [law](#) schools to see examples of programs within the same University system that made early concerted efforts to recruit and retain Native Hawaiian students, to significant success. But in fact, a focus on making the IfA a Native Hawaiian place of learning, and prioritizing the success of Native Hawaiian scholars within its halls is a recent pivot. As [Appendix N of the Astro2020 decadal survey](#) pointed out, "Since astronomical first light on Maunakea 50 years ago, there have been a total of three Ph.D.s in astronomy or astrophysics awarded to Native Hawaiians," [7]. Effort to correct for this is being put forth today by the IfA. Likewise, gains in equity and inclusion in Hawai'i astronomy institutions and observatories have been underway. We intend to continue working collaboratively on this much-needed shift throughout all of our Hawai'i astronomy institutions.

The state pursued its objective for Maunakea—the development of astronomy facilities—very effectively, and in many cases at the expense of other interests that the State was obligated to protect; not just cultural rights, but ecological concerns as well. An audit in 1998 summed up the effect of this singular focus during the preceding decades: "[the State's] focus on telescope construction on Maunakea's summit propelled the site into a premier location for astronomical research. However, this emphasis was at the expense of neglecting the site's natural resources," [18]. Subsequent to that audit, the University of Hawai'i responded with a total overhaul of its stewardship practices and has since been recognized as having one of the most effective environmental resource management programs in the islands. We acknowledge the years of effort undertaken to manage Maunakea's ecology and biology with such care over the last two decades and express gratitude to the people who work each day to carry out that mission as a part of the Center for Maunakea Stewardship.

Looking back at the historical record, from the conception of the research complex in the 1960s through to the audit in 1998, one can see that the decision to award use of Maunakea lands to the University of Hawai'i was not designed to benefit the Native Hawaiian community at large. Though there have been benefits, most notably a sizeable positive impact for the state's economy and jobs in construction and operations of observatory facilities, when we focus our attention on the direct relationship of the Native Hawaiian community to astronomy and its benefits, we must acknowledge that uplifting the Native Hawaiian community simply wasn't an objective of the endeavor. The goal of the development of astronomy was to build a world-class scientific complex on Maunakea, fuel the expansion of the higher education system, and establish the prestige of the IfA in the academic world.

A clear-eyed examination of the decision-making structures that led to the land use agreements currently in place shows a troubling scarcity of Native Hawaiian representation when major decisions were made, determining how these lands should be used. On the contrary, the history of Native Hawaiian community members' resistance to these decision-making processes is well-documented.

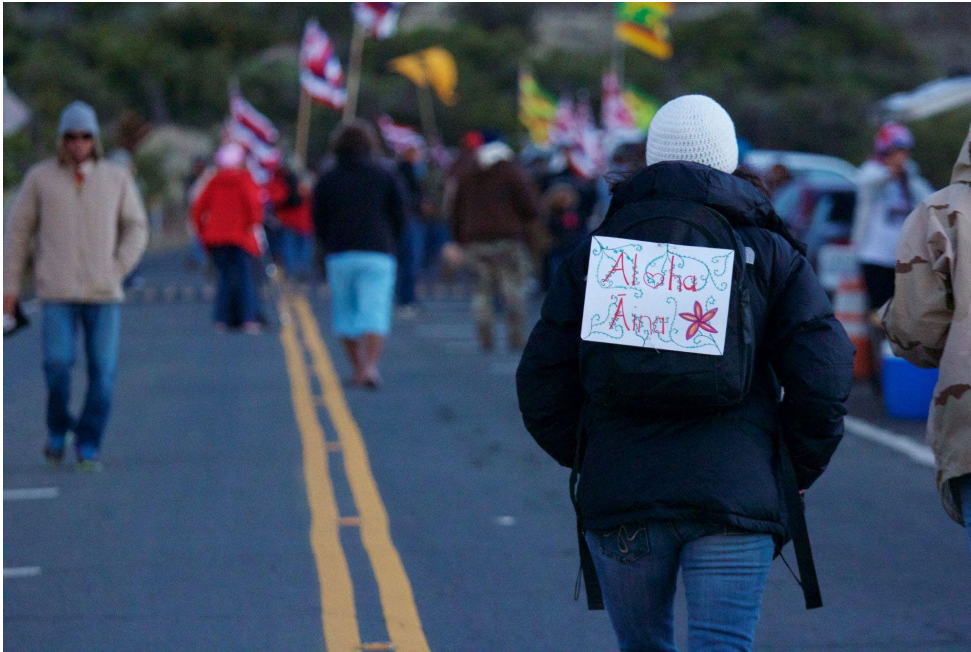

Figure 8. Activists on the Mauna Kea Access Road in 2015. The concept of Aloha ʻĀina was described by the iconic late Hawaiian environmental activist George Helm, founder of the Protect Kahoʻolawe ʻOhana: "The truth is, there is man and the environment. One does not supersede the other. The breath of man is the breath of Papa (the earth). Man is merely the caretaker of the land that maintains his life and nourishes his soul. Therefore, ʻāina is sacred. The church of life is not a building; it is the open sky, the surrounding ocean, the beautiful soil. My duty is to protect Mother Earth who gives me life. And to give thanks with humility as well as ask forgiveness for the arrogance and insensitivity of man." [19]. Photo credit: Bryson Hoe.

## 3.   THE TIPPING POINT: 5,000 KIAʻI AT PUʻU HULUHULU

**A fire ignited**

Discontent with development on the summit of Maunakea had existed for many decades, flaring into visibility at some points, but mostly out of the field of vision for the community at large and the individuals and institutions who constitute the decision-making structures that governed use of the land. But events surrounding construction at the last available site on Maunakea's summit—the largest telescope ever proposed for construction in Hawaiʻi—created a crescendo of resistance in 2019. To add context to the social, economic and political conditions that inform our view of these events see **Appendix A: "'Enough is enough' and other paths closed: a discussion of the social, economic and political conditions in which the Kiaʻi Mauna movement emerged."**

The groundswell of support for the kiaʻi, those who stood against the construction of the Thirty Meter Telescope (TMT), grew from Hawaiʻi Island to include those from throughout the continent and the world. With the legal right to build settled within the State of Hawaiʻi's court system and institutional inertia on the side of the project, there were few opposition tactics left within the conventional systems and processes, to those who were willing to give up everything to stop the TMT from being built.

Events in 2014, catalyzed by the groundbreaking ceremony planned by the TMT team and the 2015 efforts to move construction trucks to the summit to start work, sent an initial set of shockwaves through the astronomy community and local government officials, as the realization began to take hold that the status quo would no longer be a sustainable way to operate.

> *"[W]e have in many ways failed the mountain. Whether you see it from a cultural perspective or from a natural resource perspective, we have not done right by a very special place and we must act immediately to change that [.]"* Governor David Ige, May 26, 2015 [20]

Yet, it would take an even larger-scale series of events[22] to reach the scale of crisis big enough to force recognition by those in power (Observatory leadership, University of Hawaiʻi executives and directors, the Governor and his cabinet, State legislators, and Hawaiʻi's business leaders alike) that small changes would be insufficient to resolve this seemingly untenable conflict.[23] At its peak in 2019, the kiaʻi gathering at Puʻu Huluhulu would swell to the thousands on the weekends, as people clocked out of work and drove over—many even flying in from neighbor islands—to join the movement taking place there.

The State mobilized additional law enforcement officers, and TMT private security prepared to facilitate the movement of equipment; the Governor issued an emergency proclamation, enabling additional police powers and the authority to call in the National Guard.

Hawaiian environmental justice activist and attorney Dr. Trisha Kehaulani Watson reflected in *Vox*:

> *Hawaiians viewed issuance of the emergency proclamation and the growing threat of police force against them as a demonstration of how little has changed in Hawaii over the past 126 years. The unfolding events are eerily reminiscent of the events of 1893, when American and European foreigners living in Hawaii fabricated safety concerns so American military forces could land on the sovereign kingdom of Hawaii soil and aid in the coup that would allow a "union with the United States."* Trisha Kehaulani Watson, Ph.D., Why Native Hawaiians are fighting to protect Maunakea from a telescope, VOX [21]

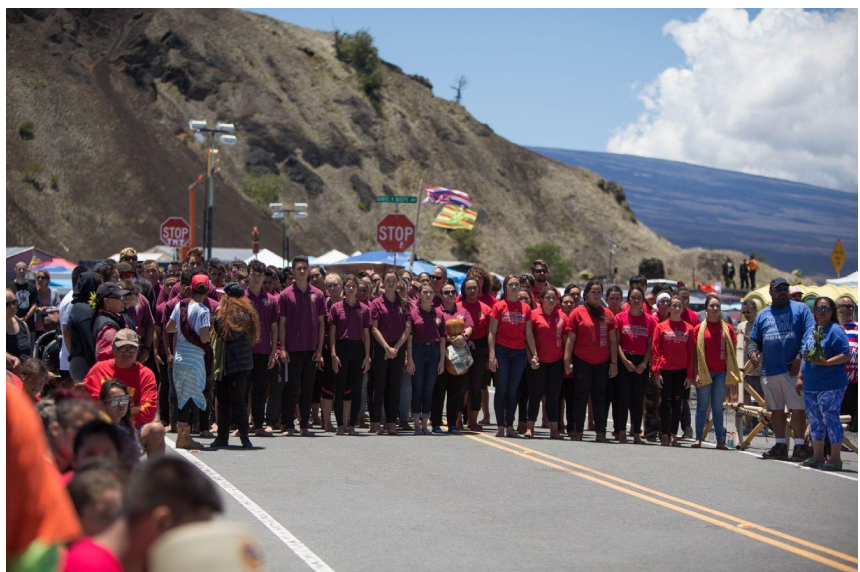

Figure 9. Students from Hawaiian-medium (language immersion) schools stand shoulder to shoulder, surrounded by community members living at the kiaʻi encampment at the base of the Maunakea Access Road during the months-long protest. Photo credit: Naʻalehu Anthony.

On the same day as the emergency proclamation, 38 kūpuna, respected elders, put themselves forth to be arrested by local law enforcement officers on the Maunakea Access Road. This was a catalytic moment that jarred the broader public into a different kind of awareness. Our community began to realize that the collective values that hold society together in Hawaiʻi were calling us to look for a way out of this seemingly intractable conflict.

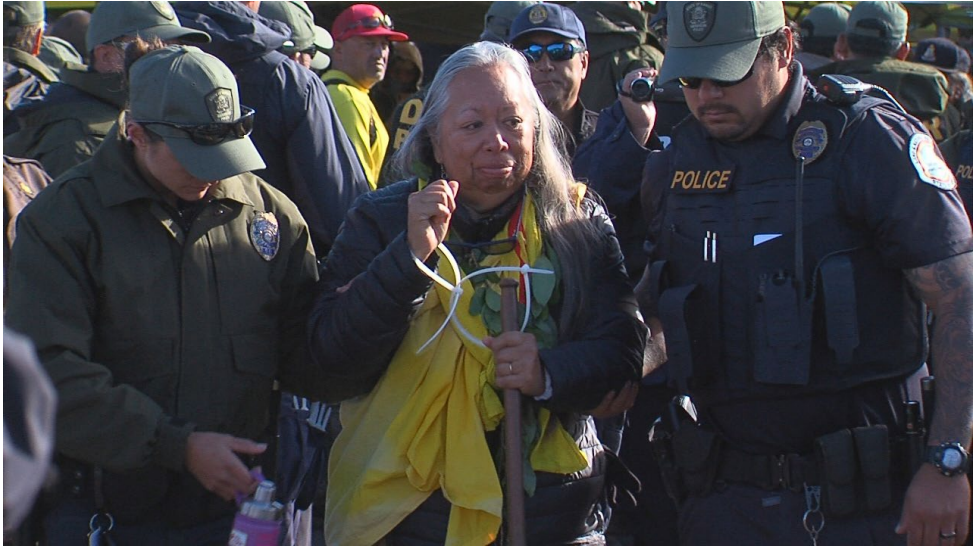

Figure 10. Noe Noe Wong-Wilson, Ph.D., executive director of the Lālākea Foundation, is one of 38 elders arrested by local law enforcement in July 2019. The arrests were captured on video by ʻŌiwi TV and can be viewed here: Maunakea Kūpuna Arrests. Today, Dr. Wong-Wilson serves as vice-chair of the newly-formed Maunakea Stewardship and Oversight Authority, working collaboratively to govern Maunakea based on mutual stewardship principles. Photo credit: Lālākea Foundation.

Within the halls of Hawaiʻi's astronomy institutions, staff members were deeply troubled and profoundly impacted by these events. Many Maunakea Observatories employees were born and raised in the islands, or are deeply integrated into the fabric of the community. Some trace their Native Hawaiian ancestry back to cultural practitioners whose traditions hold the mauna at the center. Staff members' families include those who name themselves kiaʻi, and dear friends who joined the hundreds, sometimes thousands, of community members living at Puʻu Huluhulu, the encampment at the base of the Maunakea Access Road. In these families, and in the daily lives of nearly everyone in the local community, the two parallel realities had collided. Alongside those staff members were their colleagues and friends who struggled to reconcile the purity of their intent—the pursuit of knowledge for the sake of intellectual curiosity and global advancement—with the feelings of isolation, blame, and alienation that occurred when their understanding of their identity and their place in society was fundamentally challenged. The history of disenfranchisement, theft of land, sovereignty, and culture, and the movement to reclaim identity and agency, could no longer be held apart as a co-existing parallel universe from the one in which lawful permitting of a celebrated construction project on State land might proceed. In many ways, large and small, the staff of the observatories were at the epicenter of this paradox, and the scars that resulted from holding those incongruous tensions will be long-felt. The open letter, issued by Maunakea Observatory directors in 2019, is a snapshot of the emotional turmoil and shifting perspectives experienced at that time. This was one of many opinions offered by voices in the fields of astronomy and STEM more broadly, who joined in the global discourse surrounding the conflict as it grew in intensity.[24]

In small rooms and large halls, leaders in academia, government, the business sector, and activist groups alike began to realize that a fundamental shift in *governance* would be required. The shift would need to be deeper than a change to how programs are administered, or what interests are prioritized. A fundamental change in how decisions would be made—and importantly, by whom—was the only way forward.

## 4. A PIVOT TOWARD MUTUAL STEWARDSHIP

**Introducing a mutuality framework to the Maunakea conflict**
As the COVID-19 global pandemic health crisis unfolded in the spring of 2020, a transformation of the governance and stewardship framework for Maunakea began to emerge. It is quite possible that without the pandemic-imposed timeout for activism on Maunakea, there would not have been the space for a pivot in what seemed like an intractable situation. What emerged nearly three years hence was a profound policy reform towards community-based governance and stewardship of Maunakea based on a new paradigm that came to be called "mutual stewardship."

The concept of mutual stewardship recognizes that groups of people with distinct interests in a shared land base such as Maunakea are interdependent; their futures are intertwined. According to Norma Wong, a Zen master, strategist, and policy-maker from Hawaiʻi, the aspiration of mutual stewardship is "the humans of a given place would, in relationship to each other, care for and support that place so that the place can care for and support the humans. Mutual relationship describes the relationship between the differentiated humans as much as it describes the relationship between all of the humans in a given place with that place," [22]. As such, stewardship actions and decisions must consider the whole, and not just narrow interests. Mutual stewardship requires mutual understanding between groups with differing interests, who, it should be noted, may have very different worldviews and relationships with the land. To share in the commitment to the collective thriving of the people and the land is mutual stewardship.[25]

Mutual stewardship, therefore, is in stark contrast with the stakeholder model of competing to advance or preserve one's own interests or stake, often in a transactional way, for the use of a limited resource, often resulting in conflict. In mutual stewardship, it is recognized that major unresolved conflict is a detriment to the whole.

**Policy foundations for profound reform**
There were several steps that led to major policy reform for Maunakea between 2020 and 2023. Some occurred in parallel, so it was not a linear process, but all the steps summarized below contributed to the pivot toward a new paradigm of governance for Maunakea.

In 2020 the Department of Land and Natural Resources ordered an independent evaluation of the University of Hawaiʻi's compliance with the 2009 Comprehensive Management Plan for the UH-managed lands on Maunakea. The third-party evaluator, Kuʻiwalu Consulting, gathered input from 500+ individuals and organizations covering a range of interests on Maunakea. The evaluation report published in December 2020, found that UH, "is effectively managing the activities and uses on Mauna Kea," but also stated, "...there is an absence of genuine consultation with the Native Hawaiian community that has resulted in greater mistrust of UH," [23].[26]

With evaluation in hand and having witnessed the nine-month-long stalemate on the Maunakea Access Road in 2019, Hawaiʻi House Speaker Scott Saiki demanded legislative action to reorganize the governance of Maunakea. In 2021, the State House passed resolution HR33 to form a Mauna Kea Working Group (MKWG)[27] tasked with recommending a new governance model to address shortfalls highlighted in the Kuʻiwalu report. Under the public's watchful eyes, MKWG members were selected by Speaker Saiki. It is notable that the working group included widely recognizable Native Hawaiian anti-TMT activist leaders, including Lanakila Mangauil, the cultural practitioner who famously brought the 2014 TMT groundbreaking ceremony to a halt and emerged as a charismatic spokesperson and leader of the kiaʻi; Dr. Noe Noe Wong-Wilson, an educator, Hawaiian culture-based non-profit leader, a spokesperson and leader of the kiaʻi, and one of

38 kūpuna arrested for blocking the Maunakea Access Road in 2019 (pictured in Fig. 9); and, Dr. Pualani Kanakaʻole Kanahele, a cultural knowledge expert and National Heritage Fellow of the highest regard who was also among the 38 arrested kūpuna. It was significant that those who repeatedly felt their voices were not being heard came to the table [24].

Over the second half of 2021, the MKWG met regularly and often and a tenet emerged of "putting the mauna in the middle," meaning that in the circle of people with differing interests on Maunakea, the one common and primary interest was caring for the mauna, and doing so effectively would take a holistic approach. This tenet along with grounding the group in the understanding of the cultural significance of Maunakea became foundations of the MKWG's report published in January 2022, which recommended the creation of a new governance entity independent of UH and the State Board of Land and Natural Resources, with a board including Native Hawaiians with expertise in cultural knowledge and practices [25].[28]

In parallel with the MKWG's efforts, a paper titled, "Maunakea, Mauna a Wākea: A Way Forward," lead-authored by Norma Wong with other collaborators, quietly emerged and was discussed in small groups. 'A Way Forward' called for rethinking the governance of Maunakea through the Indigenous lens of the mutuality and interconnectedness of the land and the people as the path to collective thriving. From these discussions, the term "mutual stewardship" came to be part of the vernacular and found its way into legislation for a new model of Maunakea governance [26].

During the 2022 legislative session, the State of Hawaiʻi passed a bill based largely on the governance model recommended by the MKWG. In July 2022, House Bill HB2024 became law as Act 255. Act 255's purpose is stated as establishing "...the Mauna Kea stewardship and oversight authority and the governance structure contained in this chapter to protect Mauna Kea for future generations and manage the lands contained therein for the purpose of fostering a mutual stewardship paradigm in which ecology, the environment, natural resources, cultural practices, education, and science are in balance and synergy." [27].

Meanwhile, the National Academies released Astro2020 in November 2021, just prior to the introduction of HB2024. The National Academies' panel on the State of the Profession had called for a "community astronomy model," whose principles and recommendations were in strong alignment with HB2024 and influential in the legislative process - particularly the principle of 'listen and empower' (shown below). In fact, the Maunakea Observatories referenced Astro2020 in its testimony on HB2024, supporting Astro2020's call for change in how astronomy is conducted in Indigenous spaces and local communities [28].

> *Listen and empower. Make every effort to ensure all stakeholders are heard; while it may not be possible for all to have a formal say or vote in every matter, all can have a voice, and all stakeholder voices deserve to feel listened to. At the same time, a true community-based approach empowers the local community with at least partial control, even if power-sharing is not legally required; actively listening to the community means giving the community a seat at the table where decisions are made and where governance occurs.* Astro2020, p. 3-35 [7]

**The transition to the Maunakea Stewardship and Oversight Authority (MKSOA)**
The passage of Act 255 established a new governance model for Maunakea lands that have been managed by UH up until now. Land use decisions as well as management of allowable activities on Maunakea will become the responsibility of the MKSOA whose 11 voting member board is made up of representatives of the community as well as ex-officio representatives of the university and government agencies[29]. For the first time for Maunakea, there is explicit decision-making voting authority for a recognized practitioner of traditional and customary practices and a lineal descendant of a practitioner of traditional and customary practices associated with Maunakea. There is also one board position for a representative of the Maunakea Observatories[30]. Decisions that historically gave preference to the advancement of astronomy above other interests are, by design, intended to be more balanced with consideration of astronomy alongside cultural, environmental, educational, and other community interests, but not above them.

Act 255 specifies that astronomy on Maunakea is a "policy of the state," which means that the state supports astronomy on Maunakea [27]. The Act also empowers the MKSOA to specify limitations on astronomy development in its management plan, which is yet to be developed. The MKSOA is not bound to commitments made in the existing UH master plan[31], though it is important to note that UH had already set the expectation of lowering the number of telescopes from 14 (including TMT) to nine in their latest master plan.

While the new governance paradigm brings much hope for a brighter future and collective thriving of the whole, including astronomy, there are significant practical challenges with its implementation that need to be addressed.

The transfer of management and governance responsibility from UH and the DLNR to the MKSOA must be complete by July 2028 per Act 255, which is an immense challenge given the complexities of management and the deep expertise developed by UH over five decades. Act 255 stipulates that UH's master lease with the state and the non-UH observatory subleases will continue to be valid until expiration in spite of the transfer of responsibilities. There is a moratorium in place on any new lease agreements until the transition of authority to the MKSOA is complete in 2028.

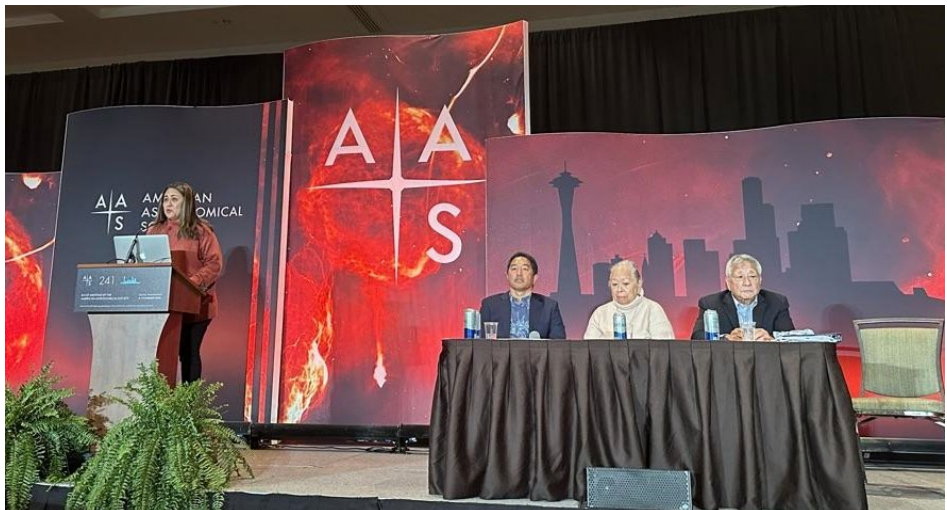

Figure 11. MKSOA Chair John Komeiji (right) and MKSOA board members Dr. Noe Noe Wong-Wilson (middle right) and Rich Matsuda (middle left) gave a plenary talk at the AAS winter meeting in 2023, moderated by Kaʻiu Kimura of the ʻImiloa Astronomy Center, addressing the profound transformation in how Maunakea will be governed moving forward. The plenary ended in a standing ovation. Photo credit: Christine Matsuda.

For the observatories, time is of the essence. The current UH master lease with the state for the science reserve and related parcels, and the subleases between UH and the non-UH observatories, all expire in 2033 [29]. This leaves just five years from when the MKSOA transition is complete to the current lease expiration for new observatory leases to be fully vetted and executed. Whereas under the prior system, UH negotiated the master lease with the state Board of Land and Natural Resources, which oversees DLNR, the observatories will negotiate new leases with the MKSOA. Contested cases and subsequent litigation challenging new leases as allowed by state law are probable and will take several years to resolve. Thus, executable leases are unlikely to be available before the observatories need to make decisions about fulfilling current lease commitments to be fully decommissioned by 2033 in the absence of new leases. The willingness of the array of international funding agencies and institutions that sponsor the Maunakea Observatories to accept these risks will emerge over the next five to ten years. Resolving the conflicting legal constraints and timelines is of paramount importance.

As of May 2024, the current status of the MKSOA is that all the authority board members have been appointed[32] and confirmed, the chairperson and two vice-chairs have been selected, public authority meetings have been occurring on a regular monthly basis since January 2023, an Executive Director, Interim Administrative Services Officer, and Executive Assistant have been hired, sufficient state funds have been appropriated, and public community engagement sessions have commenced. The authority has formed joint working groups with UH to address co-management and transition requirements. The focus has been on building organizational and administrative capacity so the authority can begin to take on substantive issues regarding the future management plan and transfer of responsibility from UH to the MKSOA. Progress is being made, and yet, the clock is ticking down.

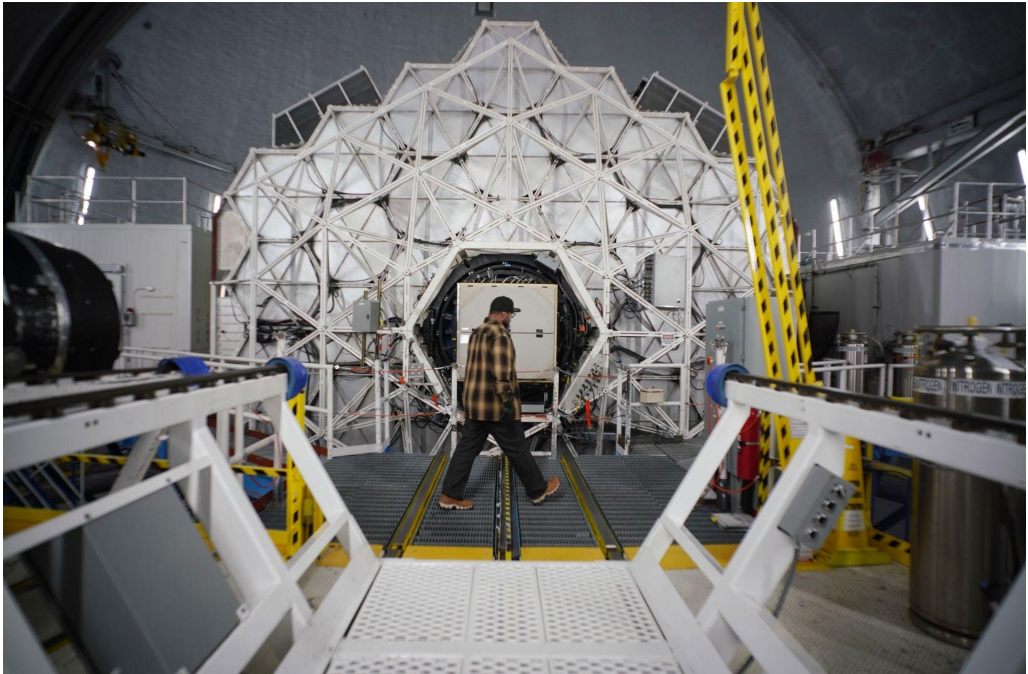
Figure 12. Day crew working at W. M. Keck Observatory in 2024. Photo Credit: Cory Lum.

**The view from the Maunakea Observatories**
An excerpt from the Maunakea Observatories' (MKO) joint comments to the Hawai'i State Legislature summarizes the shared sentiment of our leadership at that time:

> *We seek a community-based mutual stewardship model which will allow astronomy to thrive alongside other interests, sensitive to the needs of the local community. We share the attached core principles for future management of Maunakea that we believe will enable our observatories to be successful and to continue to contribute meaningfully to the people of Hawai'i and the world. We feel a mutual stewardship model reflects and is strongly aligned with the astronomy profession's broader aspirations. The recently published Astro2020 Decadal Survey, a strategic document commissioned by the National Academies of Sciences, Engineering and Medicine that provides a roadmap of astronomy priorities over the next 10 years, informs Congress and federal*

> *agencies, and presents a powerful vision for a community model of astronomy.* Maunakea Observatories, [Input to the Conference Committee re: HB2024 HD1-SD2 Relating to Mauna Kea](#) [30]

Resonance of the activism in 2015 and 2019 for the protection of Maunakea extended far beyond the island of Hawaiʻi, and shook the astronomical community as a whole. This included the Maunakea Observatories, who sat geographically at the very center of the site of the controversy. In parallel with the evolution of the political landscape with the creation of MKSOA, the Maunakea Observatories initiated a reflection to propose a model of community astronomy as it could be envisaged in Hawaiʻi. It is important to emphasize and reiterate that, whatever model we propose, the decision regarding the future of astronomy on Maunakea will be made by the MKSOA, and the observatories are collectively committed to respecting and accepting those decisions.

## 5. THE MAUNAKEA OBSERVATORIES NEW FUTURE HORIZON

Accompanying the evolution of the political landscape with the creation of MKSOA and the paradigm shift this represents, the Maunakea Observatories have initiated an effort to propose a model of community astronomy. The first elements of this model are intended to serve as an entry point for discussions with the community, and ultimately to enable the development of a co-designed and co-defined model, in full collaboration with the community, that would encapsulate the values of the observatories, the values of MKSOA, and more generally, the values of Hawaiian communities.

The first step towards this goal was to develop a horizon—a vision—of the future of astronomy at Maunakea and in Hawaiʻi. Unsurprisingly, we took a standard business approach, with a 20-year horizon, typical of our project's development timescale. The discussion was quickly hampered by practical, financial, and legal issues; for example, how to take proactive, non-discriminatory steps to modify our hiring practices to make them more inclusive of the local workforce. This first discussion had limited success. A change of methodology was called for.

One of our local advisors and facilitators, and mentor to some of us, a person with deep knowledge and experience with Indigenous and political issues, urged us to use a different approach reflective of Indigenous wisdom: the Seventh Generation Principle. This principle is a fundamental value of the Haudenosaunee (Iroquois) people and postulates that in all the decisions and actions we take today, we should consider their impact on the seventh generation to come, taking into account the seven previous generations. Compared to the original Western approach, this concept brought the immense added value of completely dissociating practical, financial, administrative, and legal issues from our vision of the future. A consensus could be reached very easily, in the form of a "Horizon Story" told by the future local leaders of Maunakea's observatories. The story articulates the hoped-for future state, one where the mauna (Maunakea) is thriving, scientifically, culturally, ecologically, and in mutuality with the community—in a word, holistically. For many of the authors of this paper, reaching a consensus on our shared vision of our future using Indigenous principles and wisdom was a first, and very significant achievement. Modest as this achievement may seem to many, we firmly believe that it illustrates how Indigenous knowledge and wisdom can and must find a place in the conduct of our profession.

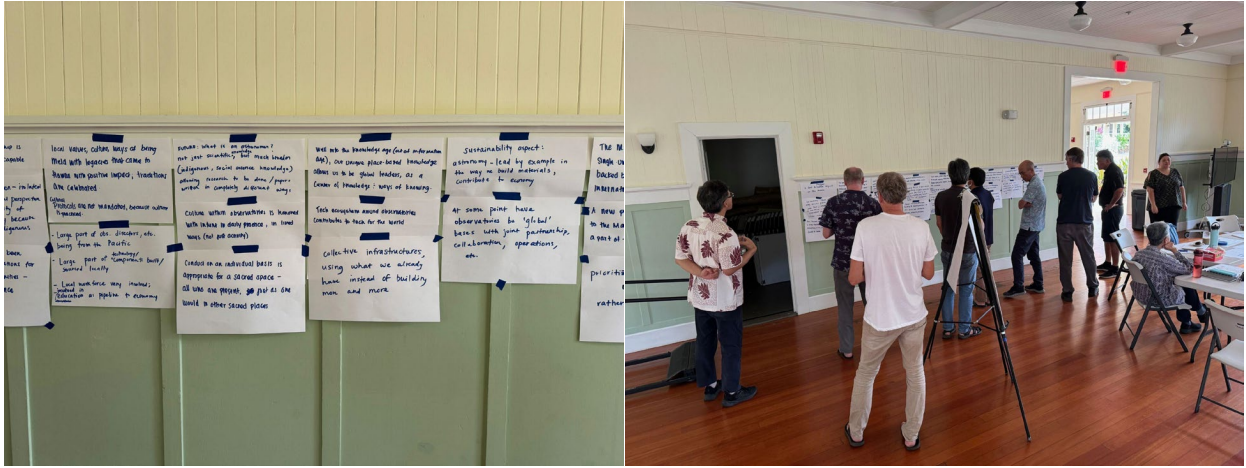

Figures 13 & 14. Maunakea Observatories directors met to develop a Horizon Story, to articulate their vision for a desirable future state. Photo credit: Christine Matsuda

The next step was to translate this story into guiding principles that should guide our actions, starting today and with the aim of making our Horizon Story a reality, generations into the future. They are grounded in four essential ideas; first, an affirmation of respect for Maunakea, which we should demonstrate through mutual responsibility in stewardship of this place. Second, that the exchange of knowledge is a central tenet of our work, not only in the pursuit of scientific discovery, but also in appreciation of the many systems of knowledge that can inform and inspire one another. Third, prioritization of our relationships with the land and community of Maunakea, which are fundamentally interconnected; to respect and care for one is to respect and care for the other, and cannot be held as separate. Fourth, a commitment to align our values and conduct to be in harmony with Hawai'i, building greater reciprocity as we operate within a global context from a place of solid local grounding in our community.

We seek to make these principles evident by living them, with the understanding that their value and integrity are defined and measured not only by words but by the actions that will follow these commitments.

As we continue to work and develop these guiding principles, they will aim to express our unwavering commitment to respect the mauna that we hold with reverence; to learn from and be inspired by other knowledge systems, including Indigenous knowledge; to nurture our connection to the land and the people who inhabit it, and to align our values with those of Hawaiian communities in our practice of astronomy.

The next step following these initial actions will be the drafting of an implementation plan, outlining short- and medium-term actions that could be undertaken with the ambition of paving the way for our Horizon Story to become a reality for our successors. Such an action plan can be expected to encompass proactive engagement with the local community, including our staff, extending the actions that all our observatories currently undertake; raising the level of awareness of the astronomical community as a whole; adapting our hiring practices to make them more inclusive and better targeted to the local workforce; improving our knowledge, understanding, and practice of Hawaiian and Indigenous culture in our day-to-day practice of the profession; developing workforce development opportunities for local schools, colleges, and universities; expanding opportunities for community members to inform and participate in our governance and to hold the highest leadership roles in our observatories.

Co-designing with the local community requires trust which is not easy to develop based on the troubled history that brought us to this point. Developing trust requires the building of genuine relations on a human-to-human level. We can only proceed at the speed of trust. We need to learn how to enter into these relationships so that trust may be built.

Furthermore, the building of these relationships cannot be solely the responsibility of a few outreach staff or thought of as an auxiliary function. Rather, community relationship building should be reflected in the core values of our organizations.

In the following, we offer proposed avenues for how we might frame community astronomy as we can envisage it in key areas involving economic benefit directly targeted to local populations, culture, environmental responsibility, and community engagement.

**Equitable hiring and workforce development:**
Observatory staff demographics should better reflect Hawaiʻi demographics, particularly in scientific, senior leadership positions. Observatory staff are community members, and these two identities should not be at odds or mutually exclusive. Unacknowledged racial and educational biases need to be addressed to remove barriers to hiring capable local applicants. Furthermore, we need to actively develop workforce development initiatives with the specific intention of creating a local pipeline for Hawaiʻi-based astronomy, engineering, and management positions. Workforce development should also serve the purpose of showing the community that astronomy is more than just astrophysicists, and does provide a wide range of jobs and opportunities. The recently funded Space Science and Engineering Initiative at the University of Hawaiʻi in Hilo will be key in enabling this endeavor, with the direct participation and support from the Maunakea Observatories.[33] In the meantime, the Maunakea Observatories have recently agreed to co-finance and establish an internship program for local students.

**Local technology development:**
All in all, the Maunakea Observatories are one of the largest concentrations of high-tech astronomy facilities in the world. Yet relatively few of the technologies they use are developed locally. Many instruments and telescope systems are imported from the institutes or countries funding the observatories. Again, the aforementioned Space Science and Engineering Initiative at the University of Hawaiʻi at Hilo will provide a tremendous opportunity to develop local technological expertise in targeted areas that can benefit the observatories and other scientific disciplines in Hawaiʻi and beyond. Our vision encompasses a hub of small- and medium-sized enterprises and start-ups clustered around the observatories and providing technologies and services to them. Pilot actions will be carried out by certain observatories to initiate the development of local expertise, in partnership with the University of Hawaiʻi, for their new instruments; some National Science Foundation programs and grants are encouraging developments in this direction.

**Cultural awareness and competence:**
Cultural orientation and education need to be broad and ongoing. Those who benefit from Maunakea, which is the majority of the world's astronomical community, along with all Maunakea Observatories staff and boards need to participate in education on Maunakea and Hawaiian culture. Hawaiian values and language, as a vehicle of culture, should be integrated into decision-making, operations, and workplace culture. Maunakea Observatories staff, and more broadly the astronomical community, should have regular opportunities to receive training and engage with cultural practitioners and community leaders to develop a deep and personal connection to Hawaiian and local culture. The Maunakea Comprehensive Management Plan from the University of Hawaiʻi currently in effect includes a requirement of orientation for all observatory staff and, more generally, all users of Maunakea services provided by the Center for Maunakea Stewardship (such as tour operators, contractors, vendors, rangers, etc.). Some observatories will extend this requirement, initially on a voluntary basis, to their user community when their proposals are accepted or when they access their data.

**Environmental responsibility:**

Sustainability and eco-responsibility are universal values that resonate strongly in Hawaiʻi and at Maunakea in particular. Astronomy at Maunakea must lead by example. The State of Hawaiʻi has committed to a full [transition away from fossil fuels by 2045](#), and the Maunakea Observatories have long been proactive in supporting this goal, installing solar panels at their headquarters and, where possible, at the summit, to partially offset their carbon footprint. The observatories have long hosted environmental monitoring equipment, which should continue to expand both in capability and interdisciplinary collaboration. New developments on Maunakea, whether instrument or telescope upgrades, should adhere to environmental best practices and minimize their ecological footprint, whether civil engineering or technological developments.

Environmental responsibility also needs to be approached from a local and cultural context, and seen as an extension of cultural awareness and competence. The Maunakea Observatories acknowledge the traditional caretakers of the spaces they use. Observatories' staff actively participate in and contribute to eco-responsible projects at Maunakea and in other parts of the island, such as removal of invasive plants and native forest restoration.

**Community engagement:**
Our traditional outreach isn't enough to develop the reciprocal relationships required for a true community-based model of astronomy. We need to reach beyond the sub-demographic of community members already engaged with astronomy, to the community at large. To achieve this, we must deeply listen and respond to the wants and needs of our community, including those who do not necessarily agree with us or support astronomy. We need to be willing to think beyond STEM and contribute to the real needs that are present. Moving from a transactional to a relationship-based mode of engagement will allow for the strengthening of reciprocal relationships that will lead to ongoing, genuine engagement that creates real value rather than performative engagement.

**Expanded scientific approach:**
The roots of astronomy lie in Indigenous culture, one of the most profound ways we can demonstrate our understanding and commitment to mutual stewardship is to integrate Indigenous ways of thinking and knowing to inform scientific inquiry and process. For scientists, and Native Hawaiian and Indigenous scientists in particular, this represents ending the dehumanization or segmenting of self that is required to do "real science." Allowing scientists and scientific inquiry to be rooted in cultural context will not only support greater scientific advancement but will make it relevant and a point of pride for the communities in which the science takes place.

**Our commitment**
In describing our current efforts to offer a different model of astronomy on Maunakea, it is important to realize and acknowledge several factors. Firstly, all the observatories operate on behalf of their funding institutions, agencies, or member states, representing the international community of users of the Maunakea Observatories. While the Maunakea Observatories can suggest avenues for change, only the leaders of stakeholders, representing the broader astronomical community as a whole, can fully engage change to bring them to fruition. Community astronomy, as a thread of change, will only be possible if the profession as a whole embraces and commits to it.

Secondly, and just as importantly, these guiding principles and implementing actions will need to be widely discussed, informed, and inspired by local communities, in the spirit of initiating what community astronomy will and should look like in the future. Whatever the current thinking of the Maunakea Observatories directors, this is by no means a top-down decision on our part. We are learning from our past mistakes, and we sincerely believe that co-defining and co-designing the future of astronomy is the only way forward. **This is our commitment.**

# 6. CONCLUSION: ASTRONOMY'S ROLE IN CREATING A BETTER WAY FORWARD

Astronomy is the scientific discipline that has the greatest impact on the public, judging by the number of press articles relaying our discoveries. This is a blessing for our discipline, but this success also obliges us.

Here in Hawaiʻi, we have been inspired by and gratefully acknowledge the work of leaders[34] like the late Dr. Paul Coleman, whose scientific achievements and Native Hawaiian cultural perspectives reminded a generation of scholars that although all too rare, there is an important place for Hawaiian astrophysicists in academia and in community. Kalepa Baybayan, Pwo navigator and prolific educator who passed away in 2021, advocated passionately throughout his career for an appreciation of the intersectionality between Native Hawaiian knowledge systems and the scientific discovery potential of large-scale telescopes. ʻImiloa Astronomy Center, situated in Hilo and led by Kaʻiu Kimura, has provided visionary leadership and a courageously safe space for the community at large to explore the many ways in which Hawaiian cultural perspectives and Indigenous practices more broadly can flourish fruitfully intertwined with the technological capabilities of telescopes.

Major efforts have been made by many people and organizations in the profession to engage the public, develop outstanding outreach material, involve the community in pioneering citizen science projects, use astronomy as a tool to generate interest in STEM in schools, and communicate with stakeholders and politicians.[35] Locally, considerable efforts have also been made to engage with the local populations. Formal educational programs like the Maunakea Scholars program and Journey Through the Universe work with teachers and the Hawaiʻi State Department of Education, customizing the experience to meet teacher and school needs rather than adopting a "one size fits all" approach. Several of the Maunakea Observatories partner with the A Hua He Inoa (AHHI) program at the ʻImiloa Astronomy Center. In the program, Hawaiian immersion school students interning at ʻImiloa use traditional Hawaiian naming practices to name instruments or discoveries.[36] AHHI is a powerful example of what is possible when we reframe our thinking about what kinds of collaboration are possible between knowledge systems: "AHHI creates pathways in which language and culture are at the core of modern scientific practice. Indigenous communities can utilize future-focused platforms like these to develop new contexts for Indigenous knowledge and language to be applied to and "live" in to ultimately ensure the health and vitality of Indigenous communities and their identities. Efforts like AHHI ultimately help to broaden the field of astronomy while also bringing Indigenous language and knowledge to the forefront by establishing deeper cultural meaning to the scientific progress made in recent years and in years to come," [31].

Despite these initiatives that have contributed significantly to raising the importance of Hawaiian culture in our profession, there is still much to be done to recognize, understand, and value the role of these Indigenous sites and the culture of the people who inhabit them, and to cultivate mutually beneficial exchanges of ideas between scholars and knowledge systems.

We know that the social diversity of the field as a whole remains "abysmal," to borrow an adjective used in the Astro2020 report. Our relationships with local communities have not always been marked by reciprocity and mutuality. The demographics of our observatories are not always representative of local demographics. Often, we import the technology for our observatories; it is not developed locally. The benefits of our activities for local populations are not obvious to many of them, apart from a generic "economic benefit" often mentioned by the profession. The demonstrations organized in Hawaiʻi to protect Maunakea made international headlines. In short, much remains to be done.

We argue that developing a community astronomy model with strong engagement with the local populations where we observe and practice will provide the best evidence-based demonstration that our community can thrive better with increased diversity and inclusivity. This will only be possible if the profession as a whole commits to working and collaborating to implement such models, raising awareness of social justice in our profession, and respecting the guiding principles that underpin this model.

This conference is aimed at engineers and technicians, as well as astronomers. As indicated in the introduction to this paper, our astronomical knowledge owes as much to the engineers, technicians, and administrative staff of the profession as to the astronomers. Similarly, changing our model of professional practice will also require the support of all our staff. Engineering tomorrow's astronomical projects will require paradigm shifts in the way they are designed, built, and operated in a community astronomy model. How can our facilities be more energy-efficient, with less ecological impact? How can we reduce our environmental footprint? How can we repurpose sites? How can local populations, with their knowledge and experience, be involved in the development of new materials and technologies? How can we integrate local know-how and culture to weave together with our current Western scientific practices? How can we involve local populations from day one in the co-design and co-development of our projects, so that they belong to them as much as to us? How can we nurture locally the future generations of scientists, technicians, engineers, administrative, outreach, and management staff who should be an integral part of tomorrow's workforce? This is a collective challenge for all of us in the profession.

In the course of our many discussions with various professional bodies and audiences, we have noted and benefited from generally very positive feedback from the younger generation. We also call on the younger generation to take ownership of this endeavor, to express it and raise it at every astronomy conference, to become key players in tomorrow's community astronomy, imbued with Indigenous values and social justice. This model of astronomy will continue to make headlines not only for its discoveries but also for its exemplary relationships with local communities.

# NOTES

1. In this paper, we use the spelling convention for "Maunakea" followed by the University of Hawaiʻi: "The University of Hawaiʻi at Hilo College of Hawaiian Language, Ka Haka ʻUla o Keʻelikōlani, recommends one word, "Maunakea" as the proper Hawaiian usage. Ka Wai Ola (Vol. 25 No. 11) also identifies "Maunakea" as the traditional Hawaiian spelling. Maunakea is a proper noun—the name of the mountain on the Island of Hawaiʻi. This spelling is found in original Hawaiian language newspapers dating back to the late 1800s when the Hawaiian language was the medium of communication. "Mauna Kea" spelled as two words really refers to any white mountain — it is a common noun (vs. the proper noun). The "Mauna Kea" spelling is only used in this document where "Mauna Kea" is used in published or legal documents, such as the "Mauna Kea Science Reserve" [32].
2. A note on the use of Hawaiian language in this paper: We use some Hawaiian terms throughout the paper in places where there is no appropriate English equivalent, or where the use of the term is commonplace in the vernacular of our community. We choose not to italicize Hawaiian terms or translate them in parenthesis because we do not view these terms as foreign. Instead, we invite readers to utilize the authoritative online Hawaiian dictionary at https://wehewehe.org/ to learn about these terms and their uses.
3. Also in Chile [33]
4. The https://native-land.ca/ website provides an overview of lands with a presence of Indigenous peoples and languages. The World Bank indicates that "There are an estimated 476 million Indigenous Peoples worldwide. Although they make up just 6 percent of the global population, they account for about 19 percent of the extreme poor. Indigenous Peoples' life expectancy is up to 20 years lower than the life expectancy of non-Indigenous Peoples worldwide." These statistics are correlated to conditions created by colonialism and the generational disenfranchisement that is its common result- seized lands, misappropriation of natural resources, criminalization of language and culture - the list is long. At the same time, leaders at the highest levels of government and academia around the world are turning to Indigenous knowledge-keepers for solutions rooted in ancestral wisdom as the most promising solutions to today's most pressing problems. The White House Council on Environmental Quality (CEQ) and the White House Office of Science and Technology Policy (OSTP)'s government-wide guidance on recognizing and including Indigenous Knowledge in Federal research, policy, and decision-making is one such example.
5. As of this writing, two of the 13 telescopes are in the process of decommissioning - the Caltech Submillimeter Observatory and the University of Hawaiʻi teaching telescope Hōkū Kea. The deconstruction and site remediation process for both telescopes is expected to be complete by the end of 2024.
6. The 2022 nomination of Maunakea to be included in the National Register of Historic Places is an informative introduction to its history, cultural, spiritual, social, and ecological significance. Today, archeologists and conservationists take special care to preserve the integrity of burial sites, stone tools, quarries, workshop complexes, and many isolated artifacts protected on Maunakea, while contemporary cultural practitioners and lineal descendents steward the living continuity of their ancestral practices. There are hundreds of documented archeological sites located in an annulus around the summit. The traditional use of Lake Waiau's water is an important aspect of the cultural significance of the site.
7. For a deeper understanding of the diverse perspectives that value Maunakea, see the Report of the Hui Hoʻolohe from the Envision Maunakea project, released in 2018.
8. The first King of Hawaiʻi officially established a unified monarchical government in 1810. The international community recognized the Kingdom of Hawaiʻi as a sovereign nation, evidenced by treaties and conventions with the United States and others.
9. The king reserved one million acres (4,050 km$^2$) for the royal family: Crown Lands. Roughly 1.5 million acres (6,070 km$^2$) were designated as Government Lands, with the remaining given to the aliʻi (royalty, nobility) and konohiki (chief regional land stewards): Konohiki Lands. For more on this topic, consider reading Native Land and Foreign Desires: Pehea LA E Pono Ai? How Shall We Live in Harmony? by Lilikalā Kameʻeleihiwa, Who Owns the Crown Lands of Hawaiʻi by Jon M. Van Dyke, and others.
10. We encourage interested readers to learn more about this period in history by reading the infamous Blount Report, a study commissioned by President Grover Cleveland to explain to the administration what had occurred in the months leading up to January 17, 1893—the day Hawaiʻi's beloved Queen Liliʻuokalani was deposed—and the events that followed.
11. Note that annexation requires agreement via a treaty, so this moniker contributes to the continued misrepresentation of historical fact.

12. "The Hui Aloha ʻĀina for Women, the Hui Aloha ʻĀina for Men, and the Hui Kālaiʻāina, along with the queen, formed a coalition to oppose the treaty." They organized mass meetings and mass petition drives which hundreds of thousands of citizens attended. Their objective was to oppose the annexation using the same democratic means that the United States abides by. On September 6, in the mass meeting at the Palace Square, President Kaulia said, "'Aole loa kakou ka lahui e ae e hoohuiia ko kakou aina me Amerika a hiki i ke Aloha Aina hope loa' (We, the nation [lāhui] will never consent to the annexation of our land to America, down to the very last Aloha ʻĀina)," [34]. And so, they sailed. The men's branch, women's branch, and the Hui Kalaiʻāina voyaged to each island, organizing mass petition drives. A total of 38,000 signatures was collected, undeniable proof that the people of Hawaiʻi did not consent to the annexation of their home." [Hawaiʻi - Kūʻē Petitions Anti-Annexation, University of Hawaiʻi library archives]. For more information, read *Aloha Betrayed: Native Hawaiian Resistance to American Colonialism (American Encounters/Global Interactions)* by Noenoe K. Silva. Also see Letter from Liliʻuokalani, Queen of Hawaii to U.S. House of Representatives protesting U.S. assertion of ownership of Hawaii, December 19, 1898.
13. For the first half of the 20th century, land in Hawaiʻi was controlled by the United States government with the islands recognized as a territory of the United States, a result of the Organic Act of 1900. The people of Hawaiʻi did not elect Hawaiʻi's governors during that period of time but rather had governors appointed for them by the President of the United States, and confirmed by the Senate.
14. Of the original 1.8 million acres (7,300 km$^2$) of ceded lands, 400,000 acres (1,600 km$^2$) were retained by the U.S. federal government for its own uses which have included things like American military installations, bombing ranges, and bulk fuel storage. This is a parallel, but interconnected, set of issues that are controversial in our community in Hawaiʻi today.
15. The Department of Land and Natural Resources was established in Hawaiʻi Revised Statutes §26-15, which also directs the Department's administration.
16. "Gerard Kuiper and his associate Alika Herring in 1964 as a premier site for astronomical research, the summit of Maunakea steadily attracted international attention and investment for the next 50+ years. The smooth topography of this ~14,000 foot shield volcano, combined with prevailing trade wind flows from the Pacific Ocean, helps provide seeing conditions on Maunakea that are nearly unrivaled" [Doug Simons, et al, The Future of Maunakea Astronomy, 2021]. Read more about this history in A Sky Wonderful with Stars by Michael J. West from the University of Hawaiʻi Press.
17. Puʻu Poliahu is a very culturally sensitive site on the summit of Maunakea; it is the highest peak of the mountain and the home of the goddess Poliahu. From many perspectives, it was a serious transgression to have built there, which is one of many ways in which early Western astronomy activities on Maunakea did not take into account the history and knowledge of the community of generational stewards of the mountain.
18. Under General Lease No. S-4191 dated June 21, 1968, the Board of Land and Natural Resources of the State of Hawaiʻi, as lessor, issued a 65-year lease to the University of Hawaiʻi with a commencement date of January 1, 1968, and a termination date of December 31, 2033. The lease comprises roughly 11,215.554 acres (45 km$^2$), a portion of Government lands of the ahupuaʻa of Kaʻohe situated at Hamakua, Island of Hawaiʻi identified under Tax May Key: 3rd/4.4.15:09.
19. "In exchange for site access to build and operate observatories, often the governing entities for those sites receive observing time as a form of compensation for land access. Examples of this include the governments of Chile, Spain, and Hawaiʻi, all of which receive observing time on observatories they host, through their respective university systems. This allows each of these governments to support robust astronomy research and education programs at vastly reduced cost compared to building and operating entire observatory complexes themselves. In the US, this federal funding for astronomy research is akin to federal sponsorship of research in medicine/health (NIH), chemistry, geology, mathematics, etc. After a proprietary period, most of the data sourced by the MKOs is stored in public archives. This helps maximize the scientific product of and public access to this valuable resource." Doug Simons, private correspondence, 2023
20. HI Rev Stat § 171-18 states: "all funds derived from the sale or lease or other disposition of public lands shall be appropriated by the laws of the State; provided that all proceeds and income from the sale, lease, or other disposition of lands ceded to the United States by the Republic of Hawaii under the joint resolution of annexation, approved July 7, 1898 (30 Stat. 750), or acquired in exchange for lands so ceded, and returned to the State of Hawaii by virtue of section 5(b) of the Act of March 18, 1959 (73 Stat. 6), and all proceeds and income from the sale, lease or other disposition of lands retained by the United States under sections 5(c) and 5(d) of the Act and later conveyed to the State under section 5(e) shall be held as a public trust for the support of the public schools and other public educational institutions, for the betterment of the conditions of native Hawaiians as defined in the Hawaiian Homes

Commission Act, 1920, as amended, for the development of farm and home ownership on as widespread a basis as possible, for the making of public improvements, and for the provision of lands for public use." [L 1962, c 32, pt of §2; Supp, §103A-18; HRS §171-18]. Of the five uses named, the legal justification for how and why Maunakea lands can be leased for free to the university is to serve as "support of the public schools and other public educational institutions."
21. As an example, see [this excerpt from an analysis by Dr. Sai](#), submitted as an attachment to a formal letter from Maui County Councilwoman Tamara Paltin, sent to University of Hawaiʻi president Dr. David Lasner, in 2019.
22. One account of the events can be found in [Kū Kiaʻi Mauna: Historical and Ongoing Resistance to Industrial Astronomy Development on Mauna Kea, Hawaiʻi](#), Isaki, B. et al. Submitted 15 Nov 2019, Prepared for the 20 Nov 2019 Astro2020 State of the Profession and Social Impacts (SoP) panel meeting (Washington, D.C.) perspective papers pertaining to Maunakea current events contributing initial evaluations of the impact of astronomy on non-astronomy communities.
23. We invite readers to experience the pain of the conflict and the raw power of the movement through the eyes of the community itself, in addition to the outside analysis and perspectives that are often shared. A starting point may include: [Maunakea protectors rally to stop construction](#); [Sacred Hawaiian volcano is at center of centuries-old battle over Native lands](#); ['A new Hawaiian Renaissance': how a telescope protest became a movement](#); [Hawaiian Elders Protesting Telescope Construction Are Arrested](#)
24. Examples include "Reframing astronomical research through an anticolonial lens -- for TMT and beyond," [35]; "Canadian Astronomy on Maunakea: On Respecting Indigenous Rights," [36]; and "A Native Hawaiian-led summary of the current impact of constructing the Thirty Meter Telescope on Maunakea," [37].
25. For more on mutual stewardship as a conceptual framework, see "The Philosophy and Elements of Mutual Stewardship," [22].
26. The report found that UH had not effectively implemented the CMP in three areas: 1) adoption of administrative rules to manage activities to ensure protection of the resources was not timely; 2) members of the Native Hawaiian community were not adequately consulted on cultural resource issues; and, 3) the community, in particular members of the Native Hawaiian community, were not effectively engaged in education and outreach efforts, including in decision-making processes related to management of Maunakea. [UH responded to Kuʻiwalu's evaluation](#), affirming their "commitment to the collaborative stewardship of Maunakeaʻs cultural, natural, educational and scientific resources," indicating UH, too, was shifting towards a mutuality paradigm. In an opening speech for the 2021 legislature, the Hawaii Speaker of the House of Representatives Scott Saiki called for the replacement of UH as the manager of the Maunakea Science Reserve lands, stating, "Mauna Kea is a manifestation of what happens when we draw lines, work in silos and disregard different views," [38].
27. The group was made up of a chairperson, three House members, seven members of the Native Hawaiian community, and representatives from the Office of Hawaiian Affairs, the State Board of Land and Natural Resources, the UH Board of Regents, and the Mauna Kea Observatories. Rich Matsuda, the third author of this paper, was the astronomy representative in this working group.
28. The MKWG was successful in arriving at a consensus report, even when prior attempts at finding common ground during the TMT conflict did not result in tangible progress. There were several key factors that warrant further study, but to summarize for the purposes of this paper: 1) respected Native Hawaiian leaders in the MKWG provided a level of credibility in the eyes of the community; 2) the ability to meet in private allowed frank discussion; 3) dedicating significant time on relationship-building in the working group before initiating substantive discussion of issues provided a level of trust and resilience to discuss opposing views in a respectful manner; 4) dedicating significant time to understanding the cultural significance of Maunakea helped to bridge gaps through mutual understanding; 5) enacting a consensus set of grounding values and principles rooted in Hawaiian culture unified the group; and, 6) a set of culturally-grounded meeting protocols centered on mutual respect and equity of voice held the space for honest dialogue. Another important factor was that the group enlisted the help of two facilitators who moved the group forward through the work in the limited 6-month available time frame.
29. The current voting member board of MKSOA include John Komeiji, Gary Kalehua Krug, Rich Matsuda, Pomaikai Bertelmann, Joshua Lanakila Mangauil, Paul Horner, Dr. Noe Noe Wong-Wilson, Ryan Kanakaʻole, Douglass Shipman Adams, and Eugene Bal III.
30. Rich Matsuda, one of the authors of this paper, is currently occupying this astronomy representative position.
31. Also during this period, UH was deeply engaged in a multi-year effort to update the 2000 Mauna Kea Master Plan to substantially shift the university's framework for land-use decisions on Maunakea in alignment with their goals of, "responsible stewardship, maintaining leadership in astronomy, diversifying education pursuits and seeking balance among those who come to Maunakea." The updated plan, the "[Master Plan for the University of Hawai'i Maunakea Lands: E Ō I Nā Leo](#) (Listen to the Voices)," was adopted by the board of regents on January 20, 2022, before the passage of Act255. However, with the passage of Act 255 which mandates the MKSOA develop its own management plan, the future lifetime of the updated UH master plan will be cut short.
32. The members of the authority serve three-year terms and may serve a maximum of three partial or full terms. The composition of the MKSOA includes representatives with the following expertise or associations: 1) business/finance; 2) land resource management; 3) K-12 or post-secondary education; 4) Maunakea Observatories; 5) lineal descendant of a practitioner of traditional and customary practices associated with Maunakea; 6) a recognized practitioner of traditional and customary practices; 7) a nominee of the Senate President; 8) A nominee from the House Speaker; 9) an ex–officio representative of the Board of Land and Natural Resources; 10) an ex-officio representative of the Hawaii County Mayor; 11) an ex-officio representative of the UH Board of Regents; and, 12) ex-officio, non-voting Chancellor of UH-Hilo [39].
33. For more information, see [UH launching space tech development center, student-training hub](#), UH News.
34. Both of these luminaries were lost too soon and are missed tremendously by our community. Read about them and their legacies: [Chad Kalepa Baybayan, Seafarer Who Sailed Using the Stars](#); [Paul Coleman, tireless advocate for astronomy in Hawai'i](#).
35. Informal and formal education have always been the backbone of traditional community engagement programs, but the COVID pandemic opened the door to collaborations with non-astronomy, noneducational groups seeking to meet community needs; these partnerships have continued to grow and flourish post-pandemic. Observatories work with local nonprofits to participate in programs like [Kaukau 4 Keiki](#), a USDA-funded program to tackle food insecurity among children under 12 on Hawai'i Island. Many such examples exist across Hawai'i Island of community-driven programs that center the well-being of the collective.
36. To read more about A Hua He Inoa, see ['Oumuamua: First interstellar object in solar system named by UH Hilo Hawaiian language experts](#); [How an Incredible Astronomical Discovery Received Its Hawaiian Name](#); [A Hua He Inoa: Blending Traditional Hawaiian Indigenous Practices Into Astronomical Discoveries](#).

## ACKNOWLEDGEMENTS

We gratefully acknowledge Norma Wong, Ka'iu Kimura, Na'alehu Anthony, Lee-Ann Heely, and all those who reviewed this paper and offered insights that we continue to learn from today. We humbly thank all the staff of the Maunakea Observatories and all those who make up the Hawai'i Astronomy 'ohana who are the backbone of all that we do, and who give us hope and inspiration for a bright future.

# REFERENCES

1. Swanner, L.A., "Mountains of Controversy: Narrative and the Making of Contested Landscapes in Postwar American Astronomy," doctoral dissertation, Harvard University (2013); https://dash.harvard.edu/handle/1/11156816
2. Marichalar, P. "'This Mountain Is It': How Hawai'i's Mauna Kea was 'Discovered' for Astronomy (1959-79)," The Journal of Pacific History. 56(2), 119-143 (2021). https://www.tandfonline.com/doi/full/10.1080/00223344.2021.1913402
3. Konchady, T., "Mauna Kea and Modern Astronomy," Astrobites (9 November 2018); https://astrobites.org/2018/11/09/mauna-kea-and-modern-astronomy/
4. Avallone, E., "Maunakea, Western Astronomy, and Hawai'i," Astrobites (2 August 2019); https://astrobites.org/2019/08/02/maunakea-western-astronomy-and-hawaii/
5. Crabtree, D. and Simons, D. Private communications (2023-2024).
6. Sacred Land Film Project website, https://sacredland.org/ (2024). Accessed May 2024.
7. Committee for a Decadal Survey on Astronomy and Astrophysics 2020 (Astro2020) et al., "Pathways to Discovery in Astronomy and Astrophysics for the 2020s," National Academies Press. https://doi.org/10.17226/26141
8. The Kali'uokapa'akai Collective, National Register of Historic Places Registration Form (2021); Accessible online at https://drive.google.com/file/d/15dU54QrvhGDgHD9KNIBJJUedjgeTot_7/view
9. "Hawai'i History and Heritage," Smithsonian Magazine (November 2007); https://www.smithsonianmag.com/travel/hawaii-history-and-heritage-4164590/. Accessed May 2024.
10. United States Congress, Joint Apology Resolution, Statute 107, 103d (1993); Accessible online at https://drive.google.com/file/d/1IpAnObzkyI1S_2aAAHBqUtXJAdtoKkuf/view
11. An Act to Provide for the Admission of the state of Hawai'i into the Union, Act of March 18, 1959, PUB L 86-3, 73 STAT 4; https://www.doi.gov/sites/doi.gov/files/uploads/An-Act-to-Provide-for-the-Admission-of-the-State-of-Hawai.pdf
12. Jefferies, J. "Dawn of a Brilliant Opportunity," University of Hawai'i Institute for Astronomy website (2024); https://home.ifa.hawaii.edu/users/jefferies/Dawn_of_a_Brilliant_Opportunity.htm Accessed May 2024.
13. "Our History," University of Hawai'i at Mānoa website (2022); https://www.hawaii.edu/about-uh/history/ Accessed May 2024.
14. "History," Mauna Kea Astronomy Outreach Committee website (2024); https://www.mkaoc.org/history Accessed May 2024.
15. Simons, D. Private correspondence (2024).
16. Article XII, Hawaiian State Constitution. https://lrb.hawaii.gov/constitution/
17. The Office of Hawaiian Affairs vs. State of Hawai'i et al, First Circuit Court, State of Hawai'i, Civil No. 17-1-1823-11, 2027. Accessible online at https://drive.google.com/file/d/1r3rvzWth4Q_xulwiLqSORd3bVXVUekS1/view
18. Hawai'i State Auditor, Audit of the Management of Mauna Kea and the Mauna Kea Science Reserve (1998); https://files.hawaii.gov/auditor/Reports/1998/98-6.pdf
19. Helm, G., "Ka Mo'olelo o George Helm," Kū'ē Hawai'i (2024); https://kuehawaii.weebly.com/ka-mo699olelo-o-george-helm.html Accessed May 2024.
20. Ige, D. "Governor Ige's Transcribed Mauna Kea Story," State of Hawai'i Governor's Office (2015); https://dlnr.hawaii.gov/mk/files/2016/10/B.01n-Gov-Iges-10-point-plan.pdf. Accessed May 2024.
21. Watson, T.K., "Why Native Hawaiians are fighting to protect Maunakea from a telescope," VOX (24 July 2019); https://www.vox.com/identities/2019/7/24/20706930/mauna-kea-hawaii Accessed May 2024.
22. Wong, N., "The Philosophy and Elements of Mutual Stewardship," (2024); https://drive.google.com/file/d/1HGIfC6B8oXM2s4WdulxeyL8rqEC_68zv/view?usp=drivesdk Accessed June 2024.
23. Kuiwalu, "Independent Evaluation of the Implementation of the Mauna Kea Comprehensive Management Plan," (2020); https://dlnr.hawaii.gov/occl/files/2020/12/Kuiwalu-Report.pdf Accessed May 2024.
24. State of Hawai'i House of Representatives, "House Resolution No. 33: Convening a working group to develop recommendations for a governance and management structure for Mauna Kea," Hawai'i 31st Legislature (2021); https://www.capitol.hawaii.gov/sessions/session2021/bills/HR33_.HTM Accessed May 2024.

25. Mauna Kea Working Group, "He Lā Hou Kēia ma Mauna A Wākea: A New Day on Mauna A Wākea," (2022); https://www.capitol.hawaii.gov/CommitteeFiles/Special/MKWG/Document/MKWG%20Final%20Report%20.pdf Accessed May 2024.
26. Wong, N., "Maunakea, Mauna a Wākea: A Way Forward," I Ke Alo Mutual Stewardship Project, (2021); https://drive.google.com/file/d/1QiYEWfmXdWeoOmRDvUGQEpDNoGsCQamA/view?usp=sharing
27. Hawai'i State Legislature, "Act 255: Relating to Maunakea," (2022); https://www.capitol.hawaii.gov/sessions/session2022/Bills/GM1358_.PDF
28. Miyazaki, S., Lotz, J., et al. "Legislative Testimony: HB2024 HD1-SD2 Relating to Mauna Kea," (2022); https://www.keckobservatory.org/wp-content/uploads/2022/04/HB2024-HD1-SD2-Relating-to-Mauna-Kea-April-25-2022-1.pdf Accessed May 2024.
29. State of Hawai'i, General Lease No. S-1491 (1968); Available online at https://drive.google.com/file/d/1IPs9T0GQUfBy058C0l_G6oQR9e5r2VNV/view
30. Maunakea Observatories, "Input to the Conference Committee re: HB2024 HD1-SD2 Relating to Mauna Kea," (2022); https://www.keckobservatory.org/wp-content/uploads/2022/04/HB2024-HD1-SD2-Relating-to-Mauna-Kea-April-25-2022-1.pdf Accessed May 2024.
31. Kimura, K., et al., "A Hua He Inoa:Hawaiian Culture-Based Celestial Naming," Astro 2020 white paper (2021); Available online at https://drive.google.com/file/d/1QOxPV7xMFMHGmte7YcZiYF_qF_6JU05c/view
32. "Meaning of Maunakea," Center for Maunakea Stewardship website (2024); https://hilo.hawaii.edu/maunakea/culture/meaning Accessed May 2024
33. Barandiaran, J., "Reaching for the Stars? Astronomy and Growth in Chile," Minerva. 53(2), 141–164 (2015). http://www.jstor.org/stable/43548978
34. Silva, N.K., *Aloha Betrayed: Native Hawaiian Resistance to American Colonialism (American Encounters/Global Interactions)*, Duke University Press (2004); https://www.dukeupress.edu/aloha-betrayed
35. Prescod-Weinstein, C. et al., "Reframing Astronomical Research Through an Anticolonial Lens -- for TMT and beyond," Instrumentation and Methods for Astrophysics arXiv:2001.00674v1 (3 January 2020); https://arxiv.org/abs/2001.00674 Accessed May 2024.
36. Nielsen, H.R. and Lawler, S., "Canadian Astronomy on Maunakea: On Respecting Indigenous Rights," Instrumentation and Methods for Astrophysics arXiv:1910.03665v2 (8 October 2019); https://arxiv.org/abs/1910.03665 Accessed May 2024.
37. KAHANAMOKU, S. ET AL., "A NATIVE HAWAIIAN-LED SUMMARY OF THE CURRENT IMPACT OF CONSTRUCTING THE THIRTY METER TELESCOPE ON MAUNAKEA," INSTRUMENTATION AND METHODS FOR ASTROPHYSICS ARXIV:2001.00970V1 (3 JANUARY 2020); HTTPS://ARXIV.ORG/ABS/2001.00970 ACCESSED MAY 2024.
38. Lovell, B., "Saiki Wants To Take Mauna Kea Management Away From UH," Honolulu Civil Beat (2 February 2021); https://www.civilbeat.org/2021/02/saiki-wants-to-take-mauna-kea-management-away-from-uh/ Accessed May 2024.
39. Hawai'i Department of Land and Natural Resources, "Board Members," Mauna Kea Stewardship and Oversight Authority webpage (2024); https://dlnr.hawaii.gov/maunakea-authority/board-members/ Accessed May 2024.
40. "10 Questions About Mauna Kea Whose Answers Might Surprise You," Kanaeokana (2019); kanaeokana.net/10 Accessed May 2024.
41. Cabanilla, J., "UH Regents in Hilo, Jazzmin Cabanilla Testifies on Mauna Kea," Big Island Video News (17 April. 2015); https://www.youtube.com/watch?v=alLGfRCP0pM Accessed May 2024
42. Case, E., "Thirty Ways to Stand for Mauna Kea When You Cannot Physically Stand on Mauna Kea (One Way for Every Meter of the TMT)," He Wahī Pa'akai: A Package of Salt (13 July 2019); hewahipaakai.wordpress.com/2019/07/13/thirty-ways-to-stand-for-mauna-kea-when-you-cannot-physically-stand-on-maunakea-one-way-for-every-meter-of-the-tmt Accessed May 2024.
43. Feder, T., "Q&A: Kealoha Pisciotta, on Mauna Kea and Conflicts of Astronomy," Physics Today (23 Oct. 2019) physicstoday.scitation.org/do/10.1063/PT.6.4.20191023a/full Accessed May 2024.
44. "Fifty Years of Mismanaging Mauna Kea," Kanaeokana (2018); www.vimeo.com/247038723 Accessed May 2024.
45. Casumbal-Salazar, L., et al., "Enduring Hawaiian Sovereignty: Protecting the Sacred at Mauna Kea, Introduction by J. Kēhaulani Kauanui," The Abusable Past Radical History Review (2 September 2019);

www.radicalhistoryreview.org/abusablepast/forum-2-enduring-hawaiian-sovereignty-protecting-the-sacred-at-mauna-kea Accessed May 2024.
46. Goodyear-Ka'ōpua, N. "Protectors of the Future, Not Protestors of the Past: Indigenous Pacific Activism and Mauna a Wākea," South Atlantic Quarterly. Vol. 116, no. 1, pp. 184–194 (2017); https://www.researchgate.net/publication/313256643_Protectors_of_the_Future_Not_Protestors_of_the_Past_Indigenous_Pacific_Activism_and_Mauna_a_Wakea Accessed May 2024.
47. "Maunakea Archive," Hawai'i Review (2019); hawaiireview.org/mauna-kea Accessed May 2024.
48. Kanahele, T.K.H. and McGregor, D.P., "Impacts of Astronomy on Indigenous Customary and Traditional Practices As Evident at Mauna Kea," (2020); doi.org/10.6084/m9.figshare.11522289.v1 Accessed May 2024.
49. Kanuha, K., "TMT Case: Kahookahi Kanuha Final Argument." Big Island Video News (20 September 2017); https://youtu.be/8k-nM1znPso Accessed May 2024.
50. Kuwada, B.K., "We Live in the Future. Come Join Us," Ke Ka'upu Hehi Ale (7 July 2015); hehiale.com/2015/04/03/we-live-in-the-future-come-join-us Accessed May 2024.
51. Laduke, W., "Matriarch Monday: Pua Case," Indigenous Goddess Gang (30 July 2019); www.Indigenousgoddessgang.com/matriarch-monday/2019/7/29/pua-case Accessed May 2024.
52. Inouye, M. "Like a Mighty Wave: A Maunakea Film," Film directed by Mikey Inouye (2019); www.youtube.com/watch?v=4J3ZCzHMMPQ Accessed May 2024.
53. Long, K. K., "The Fight of the Indigenous Protectors of Mauna Kea," The Funambulist. no. 25 (2019); https://thefunambulist.net/magazine/25-self-defense/the-fight-of-the-Indigenous-protectors-of-mauna-kea-by-k-kamakaokailima-long Accessed May 2024.
54. Maile, U. and Wiebe, S., "States of Emergency/Emergence: Learning from Mauna Kea," Abolition Journal (10 May 2020); abolitionjournal.org/states-of-emergency-emergence-learning-from-mauna-kea-a-call-for-conversation Accessed May 2024.
55. Maly, K. and Maly, O., "Mauna Kea, Ka Piko Kaulana o Ka Aina: Mauna Kea, the Famous Summit of the Land," Kumu Pono Associates (2005); www.ulukau.org/elib/cgi-bin/library?c=mauna Accessed May 2024.
56. "Mauna Kea Series – La'akea Sanborn on Kanaka Rangers," Native Stories Podcast (29 September 2019); nativestories.org/la%CA%BBakea-sanborn-kanaka-rangers Accessed May 2024.
57. Puhipau and Landers, J., "Mauna Kea: Temple Under Siege," Nā Maka o ka 'Āina film (2005); oiwi.tv/oiwitv/mauna-kea-temple-under-siege/ Accessed May 2024.
58. Peralto, L. N., "Portrait. Mauna a Wākea: Hānau ka Mauna, the Piko of Our Ea," *A Nation Rising: Hawaiian Movements for Life, Land, and Sovereignty*, edited by Noelani Goodyear-Ka'ōpua et al., Duke UP, pp. 232–45. (2014).
59. Revilla, N., "Two Poems: Lessons in Quarantine & Maunakea," Love in the Time of COVID (25 September 2020); https://loveinthetimeofcovidchronicle.com/2020/09/25/two-poems-lessons-in-quarantine-maunakea-no'u-revilla/ Accessed May 2024.
60. "Sacred Summits," Kahea: The Hawaiian-Environmental Alliance (2020); kahea.org/issues/sacred-summits Accessed May 2024.
61. Salazar, J. A., "Multicultural Settler Colonialism and Indigenous Struggle in Hawai'i: The Politics of Astronomy on Mauna a Wākea," University of Hawai'i Mānoa PhD dissertation (2014); scholarspace.manoa.hawaii.edu/handle/10125/101135 Accessed May 2024.
62. Nogelmeier, P., "Maunakea and Maunaloa Deserve Our Respect," Honolulu Civil Beat (22 November 2017); https://www.civilbeat.org/2017/11/maunakea-and-maunaloa-deserve-our-respect/ Accessed June 2024.
63. "Guidelines for Hawaiian Geographic Names," State of Hawai'i Board of Geographic Names. v1.1 (February 2016); Accessed June 2024.

# APPENDIX A

**'Enough is enough' and other paths closed: a discussion of the social, economic and political conditions in which the Kia'i Mauna movement emerged**

In 2019—the year that saw the crescendo of the Maunakea protests—42 percent of households in the state of Hawai'i were not making ends meet; for the Native Hawaiian community on the island of Hawai'i, where Maunakea is located, that number was 59% [ALICE in the Crosscurrents, Stephanie Hoopes, Ph.D., 2023]. Even before the pandemic accelerated wild swings in the local housing market, homeownership for many local families was already a fantasy, far out of reach. Cost of living in Hawai'i is famously expensive, and when juxtaposed against the low-paying jobs most readily available in the tourism-driven economy, the result is a population stretched to the far edges of what is possible. Outward migration of the local population, including Native Hawaiians, has accelerated in the past few decades due in large part to this economic pressure [Hawai'i Perspectives, Public Resource Partnership, 2023], to the extent that the 2022 US Census found more Native Hawaiians residing outside of Hawai'i than within [U.S. Census Bureau (2012)]. The locally-rooted and Native Hawaiian families who remain in the islands do so because they actively choose this place, not because it is easy to stay.

Those who choose to remain in the islands, whatever their individual circumstances, surely share a deeper commitment to the community and to the place that means more than the sum total of the indignities, injustices and hardship that have accumulated over the past two centuries. The vibrant and powerful resurgence of Hawaiian identity and culture through practices like dance, music, non-instrument navigation and deep-sea voyaging, and growing fluency in 'Ōlelo Hawai'i as a first language, have strengthened the Hawaiian community, and continue to transform the broader community in profound ways. These core cultural practices had to be reclaimed actively, after having been systematically suppressed, and in cases even outlawed, by the territorial and eventual State government. The generation of kids that were first to be educated in the newly re-established Hawaiian medium education system of the 1970s and 1980s, grounded in their identities, genealogy, and cultural values, are charismatic and passionate leaders in some of today's prominent social and political movements, and certainly at the center of the kia'i groundswell.

Over decades, strengthening voices within the Hawaiian community had engaged with the legal processes as designed by the US government through mechanisms like public hearings, contested case law, environmental review processes, and others. But as is evident in the way these processes are structured, when it comes to development, the point of these public engagement vehicles is to provide input to a process that is already in motion, with predetermined desired outcomes. As such, the dissenting voice is sometimes appeased, sometimes acknowledged, and sometimes ignored. But in many cases, the desired outcome of the institution has so much momentum—and situated in an immovable position of power within the larger systems—that the dissent is immaterial to the project's eventual construction. The notorious exceptions to this paradigm, perhaps most notably the successful effort to halt US military use of the island of Kaho'olawe as a bombing range, were cases where grassroots leaders used unconventional, risky protest tactics outside the bounds of the 'official process' in order to be heard and effectuate real change.[1]

It was in this context that the project to build "outrigger" telescopes (in reference to Polynesian canoes) at the Keck Observatory emerged in the early 2000s. The project consisted of 6 1.8-meter telescopes, which would have surrounded the two 10-meter telescopes to operate in an interferometric mode. The project, proposed and financed by NASA, quickly ran into opposition from environmentalists and Hawaiians. Protests and legal battles lasted from 2000 to 2006, eventually leading to NASA abandoning the project. The abandonment of this project is undoubtedly the first. Unfortunately, the profession as a whole failed to grasp the full significance of the outriggers project's demise, which was a major marker of opposition to continued development at the Maunakea summit, and was an early sign of what was to come with the next major project to be proposed.

---

[1] Read more about the history and activism of the Protect Kaho'olawe 'Ohana here: Kaho'olawe: A Sacred Island